\documentclass[
 aip,
amsmath,amssymb,
 reprint,
]{revtex4-1}
\usepackage{graphicx}
\usepackage{dcolumn}
\usepackage{bm}
\usepackage{hyperref}
\usepackage{physics}
\usepackage{siunitx}
\usepackage{xcolor}
\usepackage[utf8]{inputenc}
\usepackage[T1]{fontenc}
\usepackage{natbib}
\usepackage{mathptmx}
\usepackage{etoolbox}
\usepackage[normalem]{ulem}
\usepackage{fontawesome}

\DeclareUnicodeCharacter{2033}{''}

\makeatletter
\def\@email#1#2{%
 \endgroup
 \patchcmd{\titleblock@produce}
  {\frontmatter@RRAPformat}
  {\frontmatter@RRAPformat{\produce@RRAP{*#1\href{mailto:#2}{#2}}}\frontmatter@RRAPformat}
  {}{}
}%
\makeatother
\begin{document}

\preprint{AIP/123-QED}

\title[ShearView: A Compact Stress- and Strain-Controlled Rheometer for Integrated Rheo-microscopy]{ShearView: A Compact Stress- and Strain-Controlled Rheometer for Integrated Rheo-microscopy}
\author{Nikolaos Kalafatakis}
\author{Roberto Cerbino}%
\email{roberto.cerbino@univie.ac.at}
\email{nikolaos.kalafatakis@univie.ac.at}
\affiliation{ 
Faculty of Physics, University of Vienna, Boltzmanngasse 5, 1090, Vienna, Austria 
}%

%

\date{\today}

\begin{abstract}
We present \textit{ShearView}, a compact, cost-effective, and open-source rheometer that enables both strain- and stress-controlled oscillatory shear experiments, while being fully compatible with high-resolution optical microscopy. Designed for transparency and modularity, the device integrates mechanical simplicity, dual feedback control, and real-time synchronization of rheological and optical data, thereby enabling simultaneous investigation of macroscopic mechanical response and microscopic structural dynamics across a wide range of soft matter systems. ShearView is primarily constructed from off-the-shelf components and operated via custom LabVIEW software. Calibration procedures and feedback algorithms allow for the accurate application of arbitrary stress or strain waveforms in both linear and nonlinear regimes. We validate the instrument against a commercial rheometer (Anton Paar MCR 702e), 
demonstrating excellent agreement in frequency sweeps performed in the linear viscoelastic regime and large-amplitude oscillatory shear for the materials and frequency ranges tested here. In addition, we implement non-standard rheological protocols such as chirped oscillations and recovery rheology. We further illustrate the system capabilities through synchronized imaging during echo and shear-cessation protocols, highlighting its potential to link bulk rheological response with underlying microscopic dynamics. All hardware designs, control software, and example datasets are freely available to facilitate reuse, customization, and educational deployment.
\end{abstract}

\maketitle

%

\section{\label{sec:level1}Introduction}
Understanding how soft materials respond to mechanical forces is central to both fundamental physics and technological applications. Rheology provides the quantitative framework to characterize these responses, yielding macroscopic observables such as the complex shear modulus, viscosity, and yield stress under well-defined deformation protocols. However, classical rheometric measurements average over large sample volumes and long timescales, and therefore cannot access the spatial and temporal heterogeneities that often govern the behavior of disordered systems. Phenomena such as non-affine rearrangements and dynamic heterogeneity, as well as non-idealities such as shear banding and wall slip, remain largely inaccessible to standard rheometry. Yet they are crucial for interpreting the nonlinear and time-dependent behavior of complex fluids and soft solids.

To address these limitations, an increasing number of experimental approaches combine rheological measurements with complementary techniques that probe structure and dynamics at smaller length and time scales. Among these, the integration of rheometry with optical methods—collectively referred to as rheo-optics—has proven particularly versatile. Real-space approaches such as confocal microscopy have revealed microscopic particle trajectories and non-affine displacements in sheared colloidal glasses and gels~\cite{besseling2007three}, while reciprocal-space techniques provide ensemble-averaged structural and dynamical information with high temporal resolution. Small-angle light scattering (SALS) has been used to characterize flow-induced anisotropy and ordering in complex fluids~\cite{vermant2005flow}. Dynamic light scattering (DLS) under shear has uncovered rich dynamical behavior in colloidal suspensions, including cage breaking, yielding, and the transition from reversible to irreversible dynamics~\cite{petekidis2002rearrangements,viasnoff2002rejuvenation,petekidis2008yielding,aime2019probing_I}. More recently, Differential Dynamic Microscopy (DDM), which recovers scattering-like information from real-space image sequences~\cite{cerbino2008differential,lattuada2025hitchhiker}, has been adapted to characterize particle rearrangements under shear~\cite{aime2019probing_II,villa2022quantitative}.

Various hybrid devices have been developed to combine rheology with optical access, ranging from custom shear cells to advanced instruments capable of mechanical control~\cite{petekidis2002rearrangements,wu2007new,boitte2013novel,aime2016stress,villa2022quantitative}. However, these systems often suffer from practical limitations: components may not be commercially available, technical documentation may be sparse, and instrument design may restrict flexibility. In many cases, deformation can be imposed, but the rheological response is not directly measurable. Other setups permit only one mode of control (stress or strain), or suffer from artefacts such as drift due to asymmetric actuation.

To address these challenges, we present a flexible and cost-effective rheometric platform that supports both stress- and strain-controlled operation. The system is optimized for integration with commercial inverted microscopes, but can be adapted to other experimental geometries. Its low-inertia design minimizes artefacts from mechanical acceleration, while a feedback-controlled linear actuator enables precise, programmable control over deformation. This allows not only standard oscillatory protocols, but also advanced schemes such as recovery rheology~\cite{lee2019structure,donley2020elucidating} and optimally windowed chirps~\cite{geri2018time,hudson2024sigmaowch,athanasiou2024high}, which are difficult to implement on commercial instruments.

The system is controlled entirely via custom LabVIEW software, eliminating the need for embedded firmware or proprietary interfaces, and enabling full customization of experimental sequences. Complete build documentation and control code are provided to facilitate reproducibility and adaptation.

In what follows, we describe the hardware, control system, and calibration methods in detail. We then validate the rheometer’s performance by comparing strain-controlled oscillatory tests with those obtained using a commercial Anton Paar MCR 702e. Finally, we demonstrate the utility of the setup for rheo-microscopy through a series of experiments involving echo protocols and recovery rheology, both coupled to image-based particle tracking.

\begin{figure}
   \centering
   \includegraphics[width=0.5\textwidth]{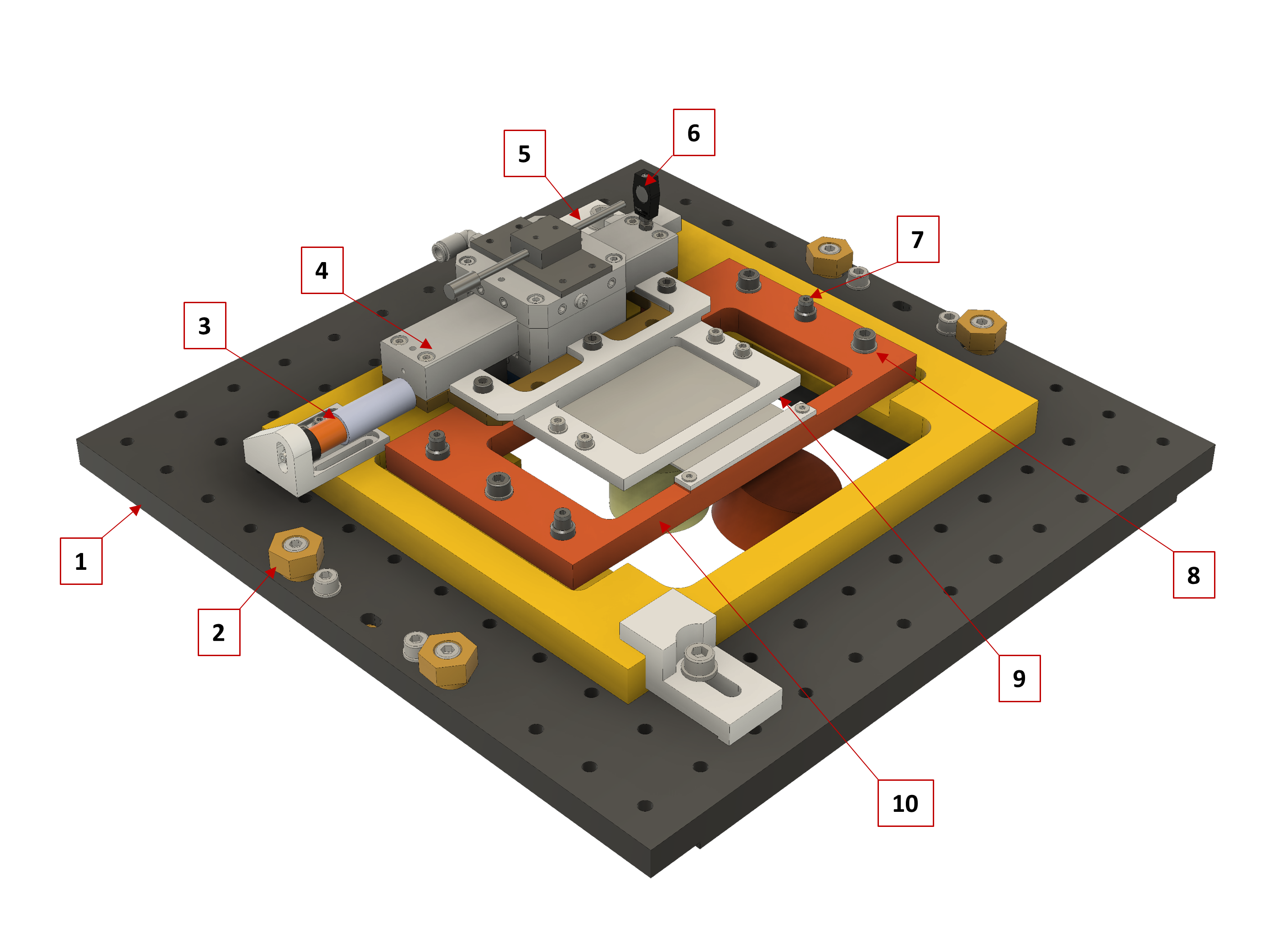}
\caption{3-D representation of the compact rheometer. 
The numbered components are:
(1) Breadboard base (MB3030/M, Thorlabs), machined to fit a commercial microscope stage;
(2) Micrometric screws, used to align the breadboard base horizontally;
(3) Linear voice-coil actuator (Moticont LVCM-013-032-02M), applying force via magnetic actuation;
(4) Linear air bearing (Physik Instrumente A-101.050), enabling frictionless translation of the moving plate;
(5) Fiber-optic sensor (Philtec D100), measuring absolute position via reflected infrared light;
(6) Gold-coated mirror (Thorlabs PF03-03-M01, mounted on Thorlabs MFM7/M), used as reflective target for the position sensor;
(7) Micrometric adjustment screws (F6MSS25 with F6MSSN1P bushings, Thorlabs) used to set the gap and fine tune the alignemnt between the two microscope slides;
(8) Locking screws, used to rigidly fix the lower structure after alignment;
(9) Upper microscope slide, mounted on the moving stage;
(10) Lower microscope slide supporting structure, fixed to the stationary base and serving as sample support.}
    \label{fig:3D}
\end{figure}

\section{\label{sec:compact}The compact rheometer}%

\subsection{\label{sec:hardware}The hardware} 

ShearView is conceptually analogous to a single-head or motor–transducer rheometer, adapted to operate in a linear shear geometry with transparent plates for optical access. A 3D rendering is shown in Fig.~\ref{fig:3D}. Full documentation, including CAD files, part numbers, and a complete bill of materials, is available on \href{https://github.com/somexlab/shearview}{GitHub} to facilitate replication and customization.

The lower plate consists of a standard microscope glass slide ($72\times52$ mm, 1 mm thick), mounted on a fixed platform. The upper plate—identical glass slide—is attached to a stage mounted on a linear air bearing (Physik Instrumente A-101.050, Germany), allowing nearly frictionless motion along the shear direction. This stage is actuated by a linear voice-coil motor (Moticont LVCM-013-032-02M, USA), which applies force via magnetic actuation. The actuator is powered by a four-quadrant Source Measure Unit (SMU, NI PXIe-4138), capable of sourcing current while simultaneously measuring voltage. Remote sensing through dedicated wires corrects for resistive losses, enabling accurate feedback directly at the actuator terminals. The force constant of the actuator—used to convert current into mechanical force and thus stress—was experimentally determined as $F = 0.911 \pm 0.002$ N/A (see Supplementary Fig.~1), consistent with previous reports~\cite{aime2016stress,villa2022quantitative}. This value is used throughout the manuscript for calculating mechanical stress from the measured current.

The gap between the plates is adjusted using three micrometric adjustment screws (F6MSS25 with F6MSSN1P bushings, Thorlabs), and parallelism is ensured through a precision-aligned base (part~10 in Fig.~\ref{fig:3D}). After alignment, the lower plate assembly is rigidly fixed with standard screws. All components are mounted on a custom-machined aluminum breadboard (MB3030/M, Thorlabs), specifically designed for compatibility with commercial microscope stages. Further details on optical integration are provided in Section~\ref{sec:synchronization}.

A distinctive feature of ShearView compared to previous implementations using low-frequency optical speckle displacement detection~\cite{aime2016stress,villa2022quantitative} is the use of a high-speed fiber-optic position sensor (Philtec D100, USA), denoted as part~5 in Fig.~\ref{fig:3D}, which brings two main advantages: i) instead of measuring the realative displacement between two times of interest, we measure the instantaneous absolute position of the moving assembly; ii) we can obtain position data (and in turn strain data) at high frequency (20 kHz), which is crucial for implementing a strain feedback loop. This sensor emits infrared light from a fiber tip and detects reflected light from a gold-coated mirror (Thorlabs PF03-03-M01, mounted on MFM7/M) rigidly affixed to the upper plate assembly. The reflected signal generates a voltage that is linearly proportional to the distance from an immobile reference surface (mirror) over a range of approximately 3.4~mm, with a sensitivity of $-0.689~\mathrm{mV/\mu m}$ in the far-side regime. The calibration curve provided by the manufacturer is reported in Supplementary Fig.~2. In addition to the standard regime, the sensor supports a high-sensitivity near-side mode for fiber--mirror distances bet ween 0.25 and 0.51~mm, with a sensitivity of $41.42~\mathrm{mV/\mu m}$. Although this mode offers increased resolution, it is limited in range and applicable only when narrow gaps or small deformations are used. Note that using a different target surface (e.g., uncoated glass or aluminum) requires adjustments to account for varying reflectivity. The sensor output is digitized using a 16-bit Data Acquisition Card (NI PXIe-6143), yielding a displacement resolution of $0.44~\mu\mathrm{m}$ over the 0--5~V range. This corresponds to a strain resolution of better than 0.1\% for typical gaps of 500~$\mu$m. For larger strain amplitudes (up to ~8 strain units), the full non-linear response of the sensor can be used at the cost of reduced accuracy.

This hardware configuration enables seamless integration with optical microscopy providing the foundation for rheomicroscopy under user-defined rheological conditions: similar to commercial rheometers, the use of a high speed position sensor allows us to complement the native stress controlled capabilities, with a feedback loop control of the strain, which represents a key advance of ShearView.

\subsection{\label{sec:software}The Software: Closing the loop of stress and strain} 

Modern rheometers typically implement control in either strain- or stress-controlled modes. In older designs, strain control was achieved by imposing a prescribed displacement via a stiff motor, with the resulting stress measured independently using a dedicated transducer~\cite{franck2008ares}. Contemporary commercial instruments are predominantly stress-controlled, where a predefined force or torque is applied by the actuator, and the resulting deformation is continuously monitored and regulated via a real-time feedback loop—often employing proportional–integral–derivative (PID) or model-predictive control schemes~\cite{lauger2002direct}.

ShearView is inherently stress-controlled: the SMU delivers a programmable current to the voice-coil actuator, which generates a mechanical force on the moving plate. This architecture has been successfully employed in previous setups combining stress-controlled rheometry with optical or scattering techniques~\cite{aime2016stress,villa2022quantitative}. In the present work, we extend this concept by implementing strain-controlled operation, achieved via real-time feedback from the high-speed displacement sensor.

The system is governed entirely by software, with all control logic implemented in LabVIEW. The position signal from the fiber-optic sensor is acquired via the analog input of the Data Acquistion (DAQ) module (NI PXIe-6143), while the control signal is output to the SMU (NI PXIe-4138), which regulates current and monitors voltage across the actuator. Both modules reside within a PXI chassis, enabling high-speed, deterministic communication.

The strain control loop operates as follows:
\begin{enumerate}
    \item The time since acquisition start is recorded.
    \item The current position of the upper plate is sampled.
    \item The target strain value at that time point is computed based on the predefined waveform.
    \item The feedback error is calculated as the difference between measured and target strain.
    \item A control current is computed using a proportional–integral (PI) control law.
    \item The computed current is sent to the actuator via the SMU.
\end{enumerate}

This control logic is executed within a LabVIEW \texttt{Timed Loop}, ensuring sub-millisecond jitter even under a non–real-time operating system. We typically employ a cycle time of 3~ms, sufficient to track sinusoidal strain waveforms of the order of 1~Hz. Supplementary Fig.~3 shows that timing jitter remains below 2~ms. Further improvements (e.g., for high-frequency operation) could be achieved by porting to LabVIEW Real-Time or FPGA-based control.

To preserve responsiveness, non-critical operations (e.g., data logging, real-time display) are offloaded to a separate \texttt{Consumer} loop in a \texttt{Producer–Consumer} architecture, ensuring that control logic remains uninterrupted.

At the start of each measurement, the actuator is automatically moved to the midpoint of its operational range, which coincides with the center of the sensor's linear response. This homing step ensures reproducibility across runs and can be optionally disabled by the user.

During acquisition, two primary signals are recorded:
\begin{itemize}
    \item The actuator current (from the SMU), converted to force and stress,
    \item The instantaneous voltage from the fiber-optic sensor, converted to position, displacement and strain.
\end{itemize}

These signals form the basis for all rheological measurements. In oscillatory experiments, storage and loss moduli are computed via Fast Fourier Transform (FFT) analysis. While this manuscript primarily reports linear viscoelastic quantities (fundamental frequency), the complete Fourier spectrum is available for LAOS and higher-harmonic analysis (see Section~\ref{subsec:rheovalidation}). In addition to sinusoidal strain, the control framework supports arbitrary target trajectories. This flexibility enables execution of complex protocols such as optimally windowed chirps~\cite{geri2018time,hudson2024sigmaowch,athanasiou2024high} and recovery rheology~\cite{lee2019structure,donley2020elucidating,lee2019recovery}, as demonstrated in Sections~\ref{subsec:chirp} and ~\ref{subsec:recovery}, respectively.

The reader interested in replicating the setup can find the control software for ShearView on \href{https://github.com/somexlab/shearview}{GitHub}. We provide, all the fundamental LabVIEW routines for basic control actions (e.g. impose current, measure position, move to home position, plot specific signals etc.) as well as one example of integration of these elementary building blocks into a fully developed implementation of a rheological test (oscillatory strain controlled shear test), using the Virtual Instrument (VI) Labview format. A screenshot of the graphical front panel interface of this VI can be found in Supplementary Fig.~4. The absence of embedded firmware and proprietary logic ensures full transparency, programmability, and reproducibility, attributes that are essential for experimental development and integration with microscopy.

In the configurations demonstrated in this work, ShearView maintained stable strain-control for sinusoidal waveforms spanning three decades in angular frequency (0.01–10~rad\,s$^{-1}$, cf.\ the chirped protocol), with accurate tracking of the target waveform as assessed by Fourier fits of stress and strain. The minimum strain amplitude we applied was set by the displacement resolution of the position sensor (0.44~\textmu m over a typical gap $h=500$~\textmu m), corresponding to strains $\gtrsim 0.05\%$. At the other end, strain amplitudes up to $\gamma_0 \approx 8$ were imposed without loss of feedback stability, limited in practice by the travel range of the moving plate. Throughout, stress-control operation remained available without modification, providing a convenient route for protocols that benefit from force actuation.

\subsection{\label{sec:analysis} Analysis of raw rheological data} 

The compact rheometer records raw position and current signals, from which strain and stress are derived in post-processing. This section outlines the analysis pipeline used to extract physical rheological quantities from these signals.

Strain $\gamma(t)$ is calculated by dividing the measured displacement $x(t)$ by the gap $h$ between the two plates:
\begin{equation}
\gamma(t) = \frac{x(t)}{h}.
\end{equation}
Stress $\sigma(t)$ is obtained by converting the measured current $I(t)$ to force $F(t)$ via the calibrated actuator constant ($F = 0.911 \pm 0.002$~N/A), and dividing by the sample contact area $A$:
\begin{equation}
\sigma(t) = \frac{F(t)}{A} = \frac{0.911 \cdot I(t)}{A}.
\end{equation}

The contact area $A$ can be determined using two methods:
\begin{enumerate}
    \item Calibration against a commercial rheometer in the linear regime, using a sample of known viscoelastic response;
    \item Direct image-based measurement, using a top-view photograph of the sample (e.g., with a smartphone) that includes a reference object of known dimensions. The sample contour is outlined, and its area computed using open-source software such as ImageJ or FIJI.
\end{enumerate}

The control loop operates with microsecond timing, but is prone to microsecond level timing variations. To enable precise spectral analysis, both stress and strain signals are resampled onto a uniform time base, using the mean sampling interval. Prior to frequency-domain analysis, a Hann window is applied to minimize spectral leakage. The Hann window at sample point $n$ is defined as:
\begin{equation}
w[n] = 0.5 \left(1 - \cos\left(\frac{2\pi n}{N-1}\right)\right),
\end{equation}
where $N$ is the total number of samples. Window-induced attenuation is corrected prior to calculating amplitude and phase.

The resampled and windowed signals are analyzed via Fast Fourier Transform (FFT), yielding the complex Fourier amplitudes of strain $\tilde{\gamma}(\omega)$ and stress $\tilde{\sigma}(\omega)$ at the driving frequency $\omega$. The complex shear modulus is then calculated as:
\begin{equation}
G^*(\omega) = \frac{\tilde{\sigma}(\omega)}{\tilde{\gamma}(\omega)} = G'(\omega) + i G''(\omega),
\end{equation}
where $G'$ and $G''$ are the storage and loss moduli, respectively.

This approach allows for high-precision extraction of viscoelastic moduli in the linear regime, and forms the basis for harmonic analysis in LAOS experiments. All scripts for signal processing (resampling, windowing, and Fourier analysis) are available in the GitHub repository.

Parameters such as the acquisition duration, sampling interval, and FFT resolution are fully user-configurable and explicitly documented to ensure reproducibility across different experimental conditions and users.

\subsection{\label{sec:inertia}Inertia corrections} 
In strain-controlled rheometry, especially at high frequencies or large strain amplitudes, inertial effects can introduce artefacts in the measured force. These arise from the acceleration of the moving components of the instrument and must be corrected to extract accurate viscoelastic moduli. In our linear shear geometry, the dominant inertial term stems from the translational mass of the moving assembly, rather than rotational inertia, relevant for commercial rheometers~\cite{franck2005understanding}.

The total actuator force $F_T(t)$, can be decomposed into two contributions:
\begin{equation}
F_T(t) = F_S(t) + F_I(t),
\end{equation}
where $F_S(t)$ is the force transmitted through the sample, and $F_I(t)$ is the inertial force associated with accelerating the moving mass $m$. The inertial force does not reflect sample properties, and must be subtracted to recover the true viscoelastic response.

Assuming harmonic motion, the inertial force is given by:
\begin{equation}
F_I(t) = -m \omega^2 x(t),
\end{equation}
where $\omega$ is the angular frequency of oscillation and $x(t)$ is the instantaneous displacement. The minus sign reflects the fact that the inertial force opposes the imposed acceleration.

\begin{figure}
\includegraphics[scale=0.4]{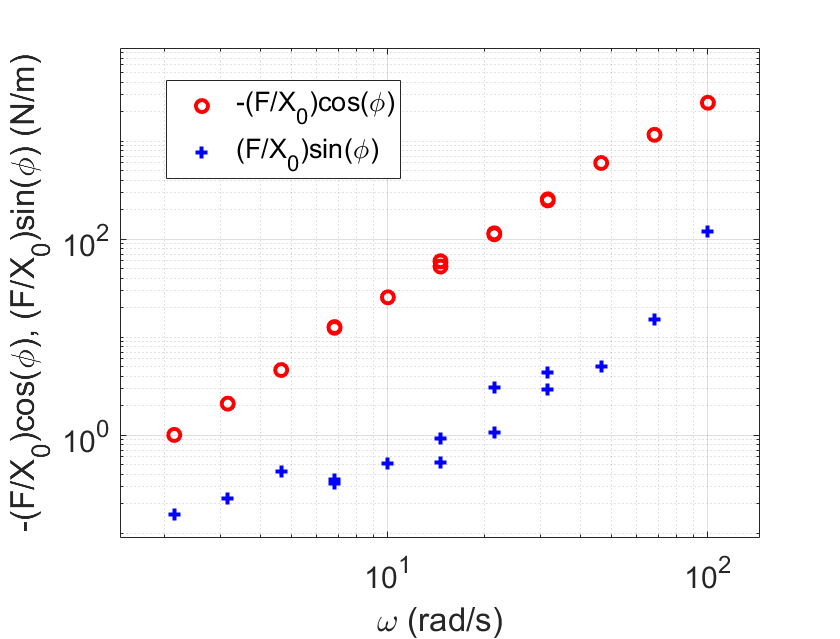}
\caption{Frequency dependence of the force amplitude normalized by displacement amplitude, measured for the empty instrument in native stress-controlled mode. The quadratic fit confirms inertial scaling, yielding an effective moving mass of $0.254$~kg.}
\label{fig:calibration1}
\end{figure}

Figure~\ref{fig:calibration1} shows the normalized force amplitude ($F_I / X_0$) as a function of frequency for the empty instrument, with $X_0$ the displacement amplitude. The observed quadratic scaling confirms the expected inertial behavior. From the fit, we estimate $m = 0.254$~kg, in excellent agreement with the independently measured mass ($0.246$~kg) of the moving parts, including the upper glass plate, actuator shaft, and mirror mount. Note that inertia measurements are performed in the native stress-controlled mode.

The inertial contribution affects primarily the storage modulus $G'$ because it is purely elastic and in-phase with strain. It does not contribute to $G''$ in the absence of damping. The corrected expressions for the moduli are:
\begin{align}
G'_S &= G'_T + G'_I = G'_T + \frac{m \omega^2 h}{A}, \\
G''_S &= G''_T,
\end{align}
where $G'_T$ and $G''_T$ are the measured moduli from the raw data, $h$ is the sample gap, and $A$ is the contact area.

The correction term
\begin{equation}
G'_I = \frac{m \omega^2 h}{A}
\end{equation}
is added to the measured $G'_T$ to recover the true storage modulus of the sample. This formulation assumes uniform stress distribution and negligible deformation of the plates. It is particularly important when working with low-viscosity or weakly elastic materials, where inertial forces may be comparable to sample response.

Notably, the need for this correction is specific to strain-controlled operation. In stress-controlled modes, the actuator compensates for inertia automatically, but at the cost of potential artefacts in $G''$ due to uncontrolled strain fluctuations. Our implementation, by contrast, combines feedback-based strain control with explicit post-hoc inertia correction, enabling accurate recovery of both $G'$ and $G''$.

Within the regimes explored in this study, the additive correction $G'_I = m \omega^2 h/A$ remained small compared to the measured $G'$ for soft materials and moderate frequencies, becoming progressively more relevant only as $\omega$ increased, in line with the expected $\omega^2$ scaling. This behavior is consistent with the independently determined effective moving mass and with standard linear-geometry analyses, and it leaves $G''$ unaffected in the absence of damping contributions from the instrument.

All processing routines for oscillatory tests (implemented in our case in MATLAB), including inertia correction, are included in the analysis scripts made available via GitHub.

\subsection{Validation of rheological performance\label{subsec:rheovalidation}}

To evaluate the quantitative accuracy of the ShearView rheometer, we performed standard oscillatory tests (namely frequency sweeps and strain sweeps) on benchmark materials representative of a broad spectrum of linear and nonlinear viscoelastic behavior. Where possible, we compared the results to those obtained on a commercial reference instrument (Anton Paar MCR 702e TwinDrive), using the same sample, to assess reproducibility and fidelity.

\begin{figure}
\includegraphics[scale=0.4]{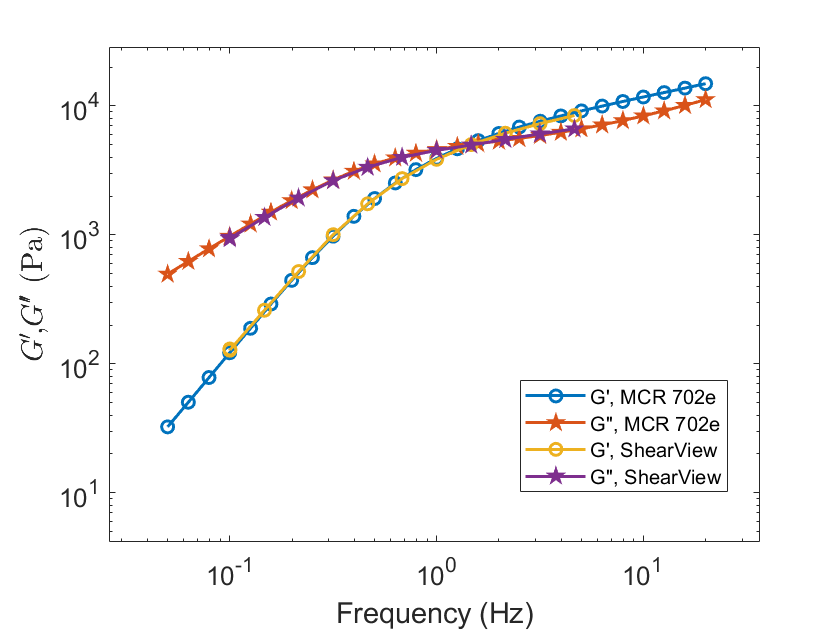}
\caption{Frequency sweep at 10\% strain amplitude for a dilute polystyrene solution. Storage ($G'$) and loss ($G''$) moduli obtained with ShearView match those from the commercial rheometer.}
\label{fig:DFS_PS}
\end{figure}

\begin{figure}
\includegraphics[scale=0.4]{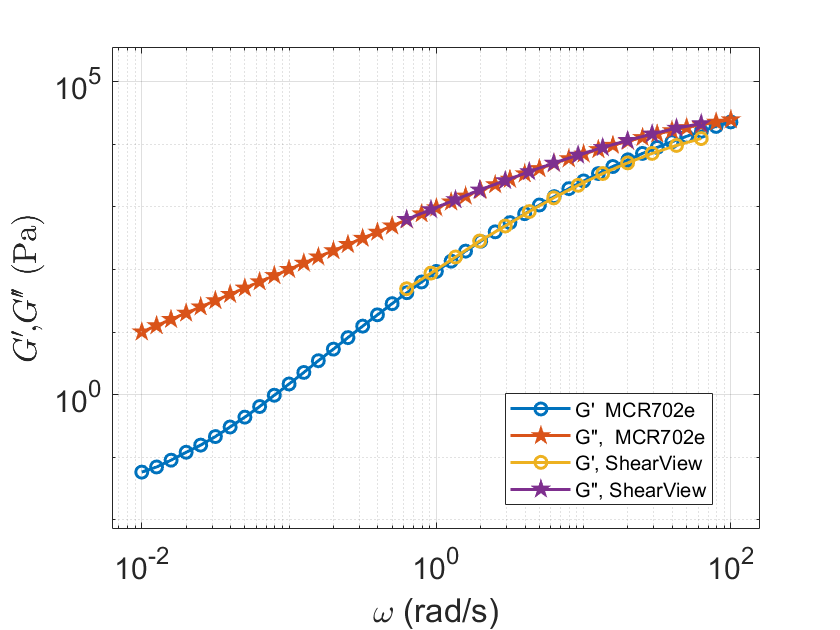}
\caption{Frequency sweep at 10\% strain amplitude for the PDMS viscosity standard, showing excellent agreement between the commercial rheometer and ShearView}
\label{fig:DFS_PDMS}
\end{figure}

Figure~\ref{fig:DFS_PS} shows a frequency sweep on a concentrated polystyrene solution, performed at a fixed strain amplitude of 10\%. The agreement with the commercial rheometer is excellent across two decades in frequency, including both the entanglement elastic plateau and the low-frequency viscous regime. The crossover point between $G'$ and $G''$ is accurately reproduced. Frequencies down to 0.1~Hz were successfully accessed, with longer acquisition times used to ensure phase stability. Similarly good data are presented in Figure~\ref{fig:DFS_PDMS}, where we report results of a frequency sweep on a high-viscosity PDMS standard.

\begin{figure}
\includegraphics[scale=0.36]{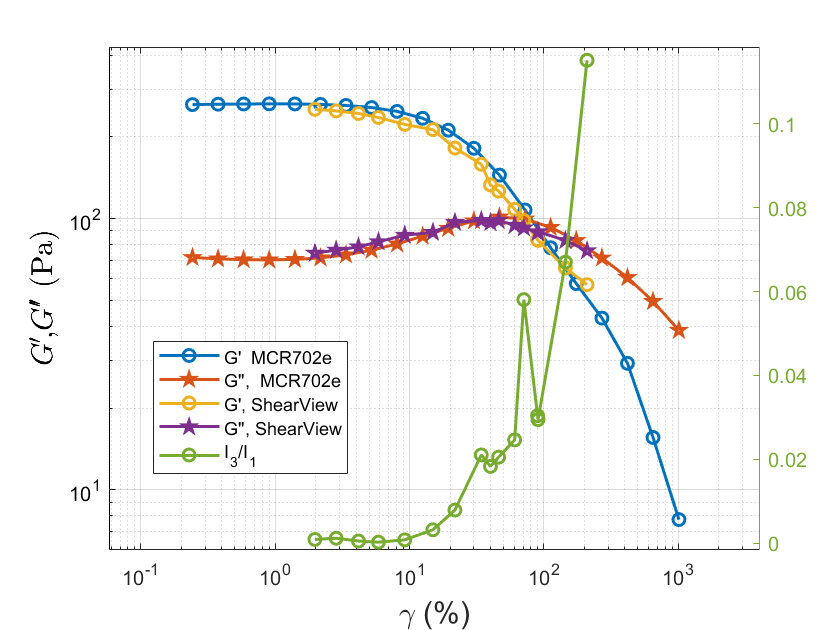}
\caption{Strain sweep at 1~Hz for the over-jammed aqueous Carbopol suspension. Right axis: normalized third harmonic of the stress signal.}
\label{fig:DSS}
\end{figure}

A complementary test of non-linear response is shown in Figure~\ref{fig:DSS}, where we performed a strain sweep at 1~Hz on an over-jammed aqueous Carbopol microgel. The device captures the key features of yielding: a linear viscoelastic regime at low strains ($G' \gg G''$), followed by a clear transition to non-linear behavior, with $G''$ overtaking $G'$ at the yield point. A small discrepancy in $G''$ (approximately 5~Pa) is observed at low strains, which we attribute to minor residual inertia, incomplete phase resolution near purely elastic response, or differences in signal filtering between instruments.

\begin{figure}
\includegraphics[scale=0.7]{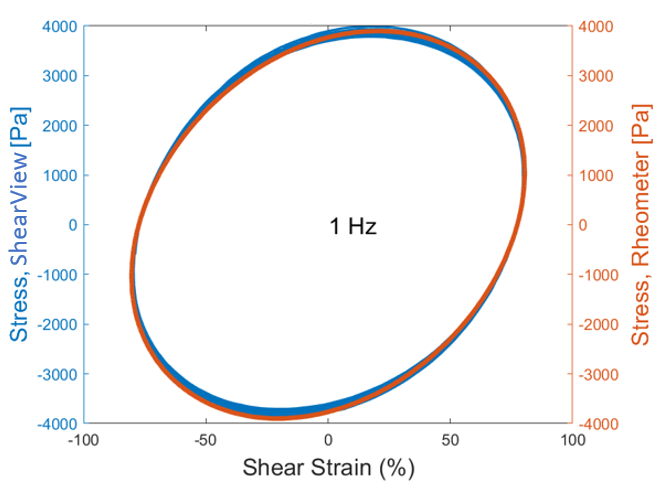}
\caption{Lissajous plot for the PDMS melt at 1~Hz. Raw signals are directly acquired and plotted without smoothing or filtering.}
\label{fig:Lissajous}
\end{figure}

Importantly, the ShearView rheometer provides access to raw time-domain data, allowing detailed signal reconstruction and harmonic analysis. Figure~\ref{fig:Lissajous} displays a Lissajous plot (stress vs.\ strain) acquired on the high-viscosity PDMS melt at 1~Hz, demonstrating the clean sinusoidal nature of both signals. No filtering or post-processing beyond FFT-based extraction was applied. This level of access is rarely possible in commercial instruments, where data are typically averaged or smoothed internally. This facilitates advanced signal processing, including harmonic decomposition and Large Amplitude Oscillatory Shear (LAOS) analysis. As an example, we show in Figure~\ref{fig:DSS} data for the ratio $I_3/I_1$ between the third harmonic and the fundamental sinusoidal component of the stress as a function of the strain.

Together, these results confirm that the ShearView rheometer delivers robust and reproducible performance in both linear and nonlinear regimes. Quantitative agreement with a high-end commercial instrument is achieved for a variety of materials, ranging from Newtonian-like fluids to yielding soft solids. The open access to raw signals, absence of embedded firmware, and full programmability of both stress and strain make the platform especially suitable for exploratory studies and custom rheological protocols.

\subsection{Flexible waveform generation: chirped strain oscillations\label{subsec:chirp}}
 
A key strength of our rheometric platform is the seamless integration of hardware and software via LabVIEW, which enables real-time control of both actuation and acquisition. This architecture removes the firmware constraints typically imposed by commercial rheometers and supports the generation of arbitrary, time-dependent stress or strain waveforms. This is an essential feature for the implementation of advanced rheological protocols, especially when coupled with microscopy.

We demonstrate this flexibility through the implementation of \textit{exponential chirp protocols}, i.e., sine waves whose frequency and amplitude evolve continuously over time. These so-called \textit{optimally windowed chirps} were introduced by Geri et al.~\cite{geri2018time} as part of the time-resolved mechanical spectroscopy method (TMS-Chirp), and have been further developed in recent work~\cite{hudson2024sigmaowch,athanasiou2024high}. Unlike classical frequency sweeps, where each frequency point is probed independently, chirps provide access to the entire frequency-dependent linear viscoelastic spectrum within a single, continuous experiment. This significantly reduces measurement time while preserving spectral resolution.

\begin{figure}
    \centering
    \includegraphics[scale=0.4]{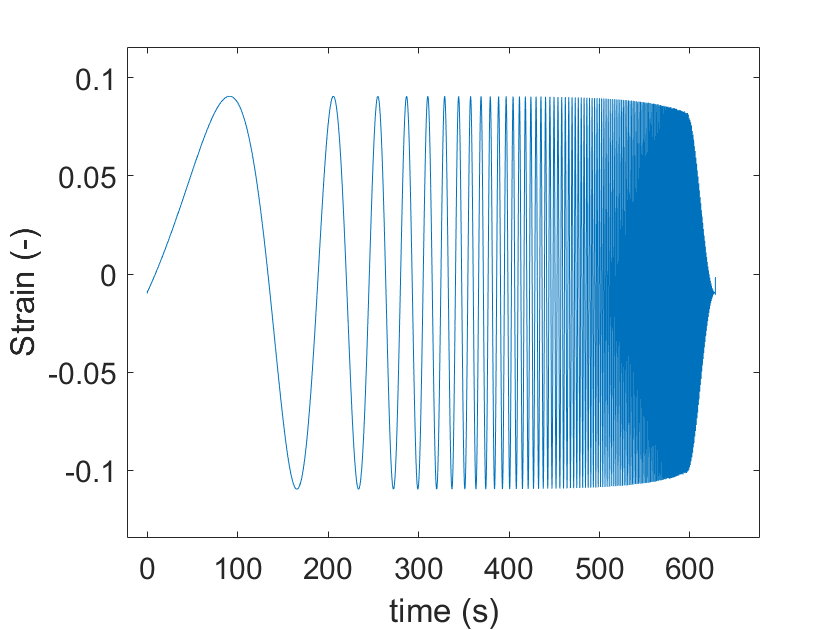}
    \caption{Strain as a function of time for an exponential strain-controlled chirp. Parameters: \(r = 10\%\), \(\omega_1 = 0.01\) rad/s, \(\omega_2 = 10\) rad/s, \(T = 628.3\) s.}
    \label{fig:strain_chirp}
\end{figure}

\begin{figure}
    \centering
    \includegraphics[scale=0.4]{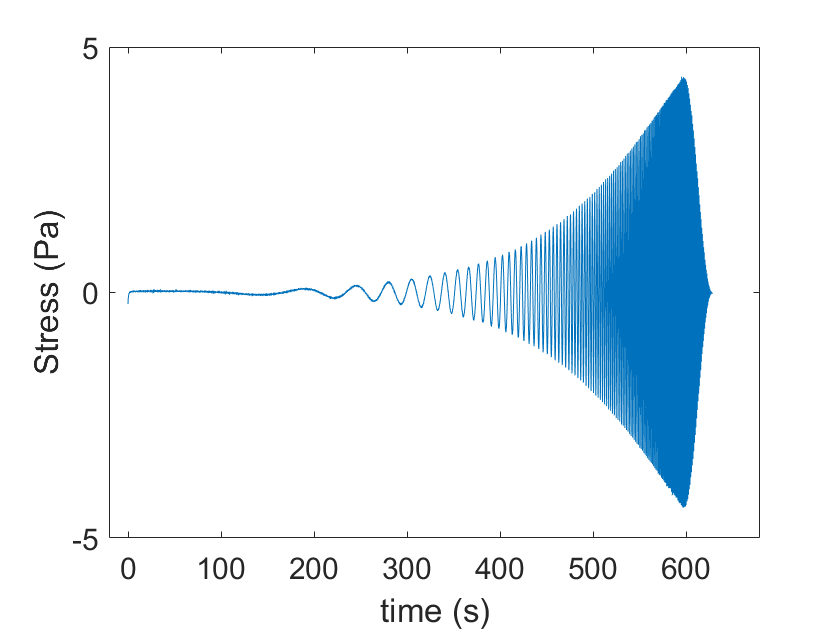}
    \caption{Stress response corresponding to the chirp shown in Fig.~\ref{fig:strain_chirp}, recorded in real time via the feedback-controlled actuator.}
    \label{fig:stress_chirp}
\end{figure}

An exponential chirp is characterized by an instantaneous angular frequency \(\omega(t)\) that increases exponentially with time:
\begin{equation}
\omega(t) = \omega_1\left( \frac{\omega_2}{\omega_1} \right)^{t/T},
\end{equation}
where \(\omega_1\) and \(\omega_2\) are the initial and final angular frequencies, and \(T\) is the total duration of the signal.

In the case of strain-controlled deformation, the strain waveform is given by:
\begin{equation}
\gamma(t) = \gamma_0 \sin\left( \frac{\omega_1 T}{\log\left( \omega_2/\omega_1 \right)} \left[ \exp\left( \log\left( \omega_2/\omega_1 \right) \frac{t}{T} \right) - 1 \right] \right).
\end{equation}

To reduce spectral leakage during Fourier analysis, we apply a \textit{Tukey window}, which smoothly modulates the signal amplitude at the beginning and end of the chirp. The window function is defined as:
\begin{equation}
w(t) =
\begin{cases}
\cos^2 \left[ \pi r \left( \frac{t}{T} - \frac{r}{2} \right) \right], & 0 \leq \frac{t}{T} \leq \frac{r}{2} \\
1, & \frac{r}{2} < \frac{t}{T} < 1 - \frac{r}{2} \\
\cos^2 \left[ \pi r \left( \frac{t}{T} - 1 + \frac{r}{2} \right) \right], & 1 - \frac{r}{2} \leq \frac{t}{T} \leq 1,
\end{cases}
\end{equation}
where \(r\) is a dimensionless tapering parameter controlling the extent of amplitude modulation.

Figures~\ref{fig:strain_chirp} and \ref{fig:stress_chirp} show a representative chirp experiment performed on a high-viscosity PDMS melt. The applied strain signal spans over three decades in frequency (from 0.01 to 10 rad/s) within a total duration of 628.3~s. The resulting stress response exhibits a rich structure reflecting the viscoelastic nature of the sample.

\begin{figure}
    \centering
    \includegraphics[scale=0.4]{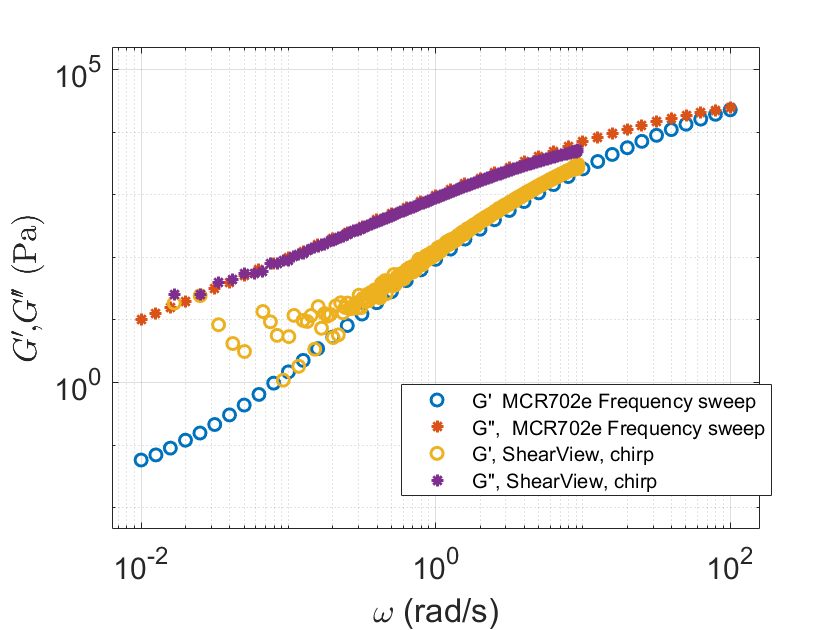}
    \caption{Comparison of dynamic moduli obtained using a chirped strain protocol and a classical frequency sweep on a PDMS melt.}
    \label{fig:chirp}
\end{figure}

To validate this approach, we compare the dynamic moduli obtained from the chirp using ShearView with those extracted from a standard Dynamic Frequency Sweep (DFS) performed on the same sample using an Anton Paar MCR 702e instrument. As shown in Fig.~\ref{fig:chirp}, the agreement between the two methods is excellent, particularly in the intermediate frequency range. Minor discrepancies at very low frequencies can be attributed to signal-to-noise limitations or transient effects during the onset of deformation. We note that similar chirp tests conducted with the rheometer using the single-motor configuration lead to data (not shown here) of lower quality, as expected due to higher flexibility of ShearView software compared to the one of commercial rheometers\footnote{One way to obtain higher quality data with the commercial rheometer would be to operate it in a separated motor-transducer configuration, which for the MCR 702e used in this work would imply using a second motor for the lower plate.}.

Importantly, the entire chirp protocol requires only about 10 minutes to complete, an order of magnitude faster than the DFS acquisition, which takes nearly two hours using standard settings (7 points per decade). This demonstrates the efficiency and utility of chirped protocols for fast, high-resolution mechanical spectroscopy.

In summary, the implementation of chirped strain oscillations on our platform illustrates the advantages of full waveform programmability and high-frequency position feedback. Such protocols enable rapid, information-rich rheological characterization and can be combined with microscopy for time-resolved studies of structural dynamics.

\subsection{Flexible waveform generation: recovery rheology\label{subsec:recovery}}

Another important example of advanced rheological protocols enabled by our system is \textit{recovery rheology}~\cite{lee2019structure,donley2020elucidating}, a technique developed to quantify the recoverable and unrecoverable components of strain upon cessation of shear. This approach is particularly useful in the study of yielding materials and allows time-resolved mechanical characterization beyond standard steady-state or oscillatory protocols. Despite its growing relevance, recovery rheology is not natively supported by commercial instruments. For instance, on the Anton Paar MCR 702e, implementation requires manual construction of strain or stress waveforms via tabulated values of time and amplitude. These must be imported as external datasets for each individual experiment, which is both time-consuming and inflexible. Moreover, raw output signals are typically processed or filtered by proprietary software, making it difficult to perform custom analyses or verify the underlying data.

By contrast, our system offers direct, programmable implementation. Both actuation (via current sourcing) and sensing (via position measurement) are governed by LabVIEW routines, and transitions between deformation phases (e.g., from oscillatory shear to zero stress) are precisely controlled within the same loop. Synchronization with optical imaging is handled through a digital trigger output that signals the camera to initiate acquisition at the precise moment shear ceases.

We demonstrate the method using an over-jammed Carbopol microgel suspended in propylene glycol. The protocol comprises three phases: i) An initial oscillatory shear of fixed amplitude and frequency applied for four full cycles, allowing the sample to reach a dynamic steady state; ii) Continuation of the oscillation for a variable fractional period (ranging from $1/32$ to $1/2$ of a cycle), to modulate the total strain accumulated before cessation; iii) Abrupt cessation of shear, implemented by setting the control current to zero. Simultaneously, a trigger pulse initiates synchronized imaging at 100 fps.
\begin{figure}
    \centering
    \includegraphics[scale=0.4]{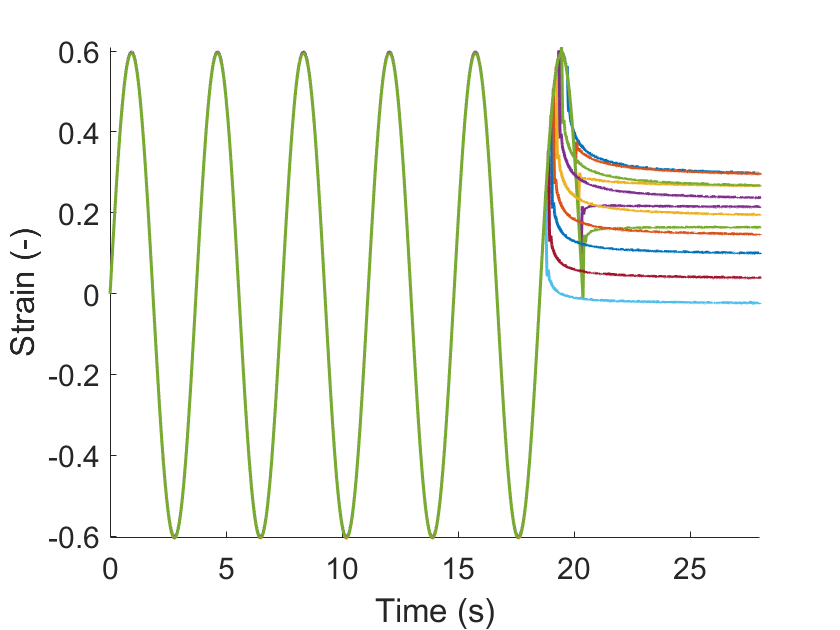}
    \caption{Representative recovery rheology experiment. The sample is subjected to 4 full cycles of oscillatory shear, followed by a fractional cycle and then abrupt cessation of stress. Strain is monitored during and after shear. Here, only one direction of shear is shown for clarity (see also Supplementary Figure 12).}
    \label{fig:recoveryForward}
\end{figure}
Figure~\ref{fig:recoveryForward} shows typical strain responses for various cessation phases. To isolate the reversible and irreversible strain components, each measurement was repeated with reversed deformation polarity. By averaging over forward and reverse runs (a representative example of the strain traces obtained in both directions is shown in Supplementary Fig.~12), we cancel minor mechanical asymmetries and obtain a clean separation of the total strain immediately before cessation into recoverable and unrecoverable contributions:
\begin{equation}
\gamma_{\mathrm{total}} = \gamma_{\mathrm{recoverable}} + \gamma_{\mathrm{unrecoverable}}.
\end{equation}
Raw signals from the device were processed using MATLAB routines based on the open-source code of Donley et al.~\cite{donley2020elucidating}, with minor adaptations to account for the specific format and scaling of our current and displacement data. Results for the strain decomposition are visualized in Fig.~\ref{fig:LissajousRecovery}. The left panel shows a standard Lissajous plot of stress versus total strain. The center and right panels depict the extended Lissajous curves where the stress at cessation is plotted against the unrecoverable and recoverable components of strain, respectively. These visualizations provide direct insight into the sample capacity to store and dissipate mechanical energy under oscillatory loading. Further analysis of these data goes beyond the scope of the present work.

\begin{figure}
    \centering
    \includegraphics[scale=0.25]{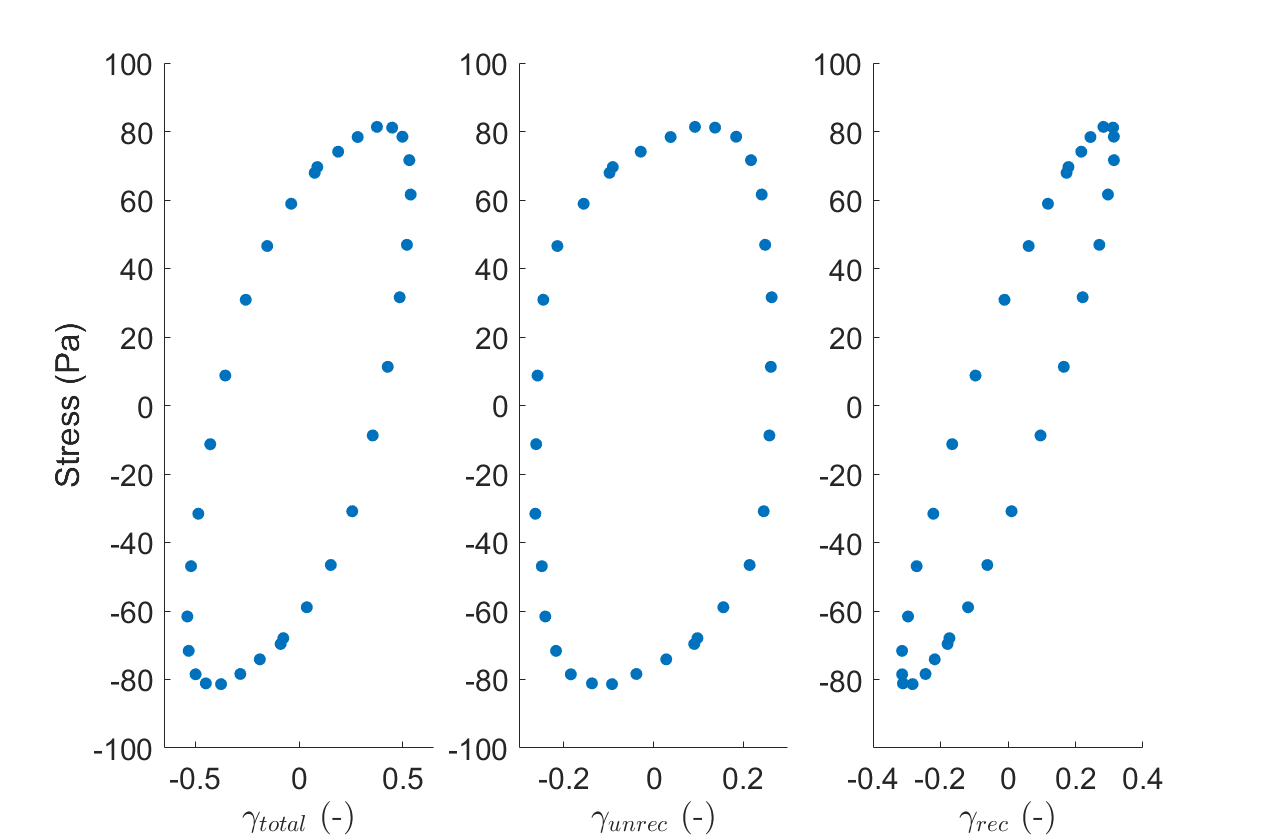}
    \caption{Traditional (left) and decomposed (center and right) Lissajous plots. Left: stress vs.\ total strain. Center: stress vs.\ unrecoverable strain. Right: stress vs.\ recoverable strain.}
    \label{fig:LissajousRecovery}
\end{figure}


In summary, recovery rheology showcases the unique capabilities of our system: programmable waveform generation, seamless control of deformation and acquisition, and precise synchronization with imaging. Together, these features enable detailed mechanistic studies of non-linear and time-dependent phenomena in soft materials, functionality that is rarely available without extensive customization in commercial instruments.

\section{\label{sec:microint} Integration with Microscopy}

\subsection{\label{sec:synchronization}Imaging synchronization} 
A key feature of ShearView is the ability to synchronize optical microscopy with rheological actuation in a reproducible and hardware-deterministic fashion. This synchronization capability applies to both periodic and arbitrary deformation protocols, enabling imaging to be initiated from a well-defined and repeatable strain state across cycles. The underlying strategy relies on monitoring the analog output of the displacement sensor and comparing it to a programmable DC voltage reference generated via the DAQ. This comparison is executed in real time using a custom-built comparator circuit based on a Schmitt trigger~\cite{melek2016analysis,pfister1992novel}, which generates a transistor–transistor logic (TTL) pulse whenever the displacement signal crosses the reference threshold. The generated TTL pulse triggers a counter module on the DAQ, which then produces a programmable train of TTL pulses controlling the timing and number of camera exposures. This mechanism is implemented entirely in hardware, avoiding latencies or jitter introduced by operating system processes.

Each time the displacement signal crosses the threshold, a new acquisition cycle is initiated, ensuring that image sequences begin at the same strain value or phase, independent of waveform complexity. This is particularly advantageous in oscillatory shear for reconstructing stroboscopic dynamics, but it also applies to non-periodic or transient protocols such as chirps and shear cessation. The DC threshold is user-adjustable, enabling imaging to start at specific strain values, and the same TTL trigger can be used to synchronize auxiliary instruments or count cycles. The delay between threshold crossing and the first camera trigger is $<30$~\textmu s, corresponding to a relative timing error below 30~ppm at 1~Hz, and remained within this bound over extended acquisitions with no measurable phase drift across repeated cycles.

\subsection{\label{sec:micro}Microscopy} 
Integration of microscopy with rheology requires samples to provide sufficient optical contrast for imaging microstructural dynamics. This contrast may be intrinsic or introduced via tracer particles, such as refractive index–mismatched or fluorescent beads. Depending on the experimental goals, different microscopy modalities can be employed, including bright-field, phase contrast, fluorescence, or confocal imaging. Camera selection similarly depends on sensitivity, bit depth, and frame rate requirements.

Our rheometer is specifically designed to be compatible with inverted optical microscopes. For this study, we mounted the apparatus on a Nikon Eclipse Ti-2U by replacing the standard microscope stage with a machined aluminum breadboard (MB3030/M, Thorlabs; component 1 in Fig.~\ref{fig:3D}). This breadboard is supported with two screws (component 2 in Fig.~\ref{fig:3D}) for coarse leveling. While optimized for Nikon instruments, this mounting approach can be adapted to other microscopes by modifying breadboard dimensions and mounting geometries.

The shearing geometry employs two standard microscope slides ($72 \times 52$~mm, 1~mm thick). The upper slide is fixed to the actuator stage, while the lower slide rests on a support adjustable via three micrometric adjustment screws (F6MSS25 with F6MSSN1P bushings, Thorlabs). The shear gap is defined optically: first, the upper slide is brought into focus using a 20$\times$ objective; then, the microscope nosepiece is lowered by the desired gap distance. The lower slide is then aligned to this position using the adjustment screws. Similar to what done in Refs.~\cite{aime2016stress,villa2022quantitative}, to achieve parallel alignment between the glass plates, an expanded converging laser beam is introduced into the illumination path and focused onto the back focal plane of a low magnification 2X objective. Interference fringes arising from reflections at the two slide surfaces are monitored and minimized through iterative adjustment of the adjustment screws. Representative fringe patterns before and after alignment are shown in Supplementary Material Fig.~6.

\begin{figure}
\includegraphics[scale=0.32]{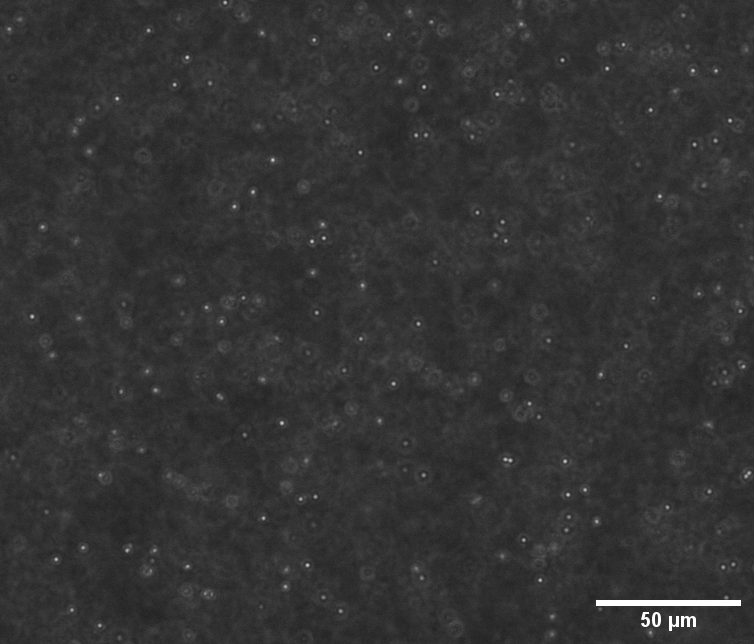}
\caption{Representative phase contrast image showing tracer particles embedded in a Carbopol gel. The sharp circular features correspond to 2~\textmu m polystyrene beads used to monitor microscopic displacements.}
\label{fig:contrast}
\end{figure}

For the experiments in this work, we used phase contrast microscopy with 2~\textmu m diameter polystyrene tracer particles (MicroParticles GmbH), dispersed at 0.05~wt\%. These particles are nearly density-matched to both water- and propylene glycol–based suspensions (see also Section \ref{sec:samples}) and exhibit minimal sedimentation. Their optical properties yield high contrast in phase contrast mode, appearing as bright circular features on a dark background (see Fig.~\ref{fig:contrast}; Supplementary Fig.~5 shows a bright-field image of the same sample). Imaging was conducted using a long working distance 20$\times$/0.45 ELWD objective, equipped with a matched phase ring and annular condenser diaphragm. Alignment was confirmed using a Bertrand lens to ensure overlap of the condenser and objective back focal planes. Images were acquired using a Ximea MC050MG-SY monochrome camera, in 8-bit format. The typical field of view was $425 \times 355$~\textmu m.

\subsection{\label{sec:micro_da} Microscopy data analysis}  
Microscopic dynamics are extracted through offline analysis of image sequences acquired during the experiment, processed using user-developed scripts (MATLAB in this work). Two complementary approaches are employed: particle tracking and image registration. These provide access to different aspects of microstructural behavior.

Particle tracking enables the reconstruction of individual tracer trajectories, revealing non-affine rearrangements, spatial strain heterogeneity, and local fluctuations. In contrast, image registration quantifies bulk or mesoscopic affine displacements by computing relative shifts between consecutive frames. Both methods are essential, depending on the experimental objective.

Image registration is typically used for reconstructing spatially resolved deformation profiles, especially in $z$-scans across the gap during oscillatory shear. Here, the relative shift between frames reflects the imposed local deformation. For tracer-based experiments such as recovery protocols, registration is not required unless global drift or stage instability interferes with tracking accuracy.

When needed, image registration is performed using the subpixel-accurate DFT-based algorithm by Guizar-Sicairos et al.~\cite{guizar2008efficient}. This method calculates translational offsets by maximizing cross-correlation in Fourier space and refines the result via inverse FFT to obtain subpixel precision. The resulting displacements are then integrated along the $z$-axis to reconstruct shear profiles (e.g., Figs.~\ref{fig:profiles_solid} and \ref{fig:profiles_nanoemulsion}).

Particle displacements are extracted using centroid-based localization. Particle centers are identified and fitted with a 2D Gaussian to achieve sub-pixel precision. Tracking is performed using modified routines from the Kilfoil Lab package~\cite{pelletier2009microrheology}, originally developed for dense colloidal systems.

Localization is followed by frame-to-frame linking using a standard nearest-neighbor algorithm with maximum displacement constraints, yielding complete particle trajectories over time. These are used to compute probability distribution functions (PDFs), mean-squared displacements, and other statistical observables that characterize microscopic response and recovery.

Unless otherwise stated, image sequences are analyzed in raw format: no temporal averaging, background subtraction, or digital filtering is applied before processing.

This analysis pipeline enables quantitative correlation between macroscopic deformation and microstructural rearrangements, under both steady-state and time-dependent shear protocols. However, what we have done here should be considered just as one of the many existing implementations to quantify affine and non-affine displacement.

\subsection{Experimental determination of shear profiles}
To verify that the compact rheometer imposes a well-controlled and spatially uniform deformation field, we measured shear profiles across the sample gap using time-resolved microscopy under oscillatory strain. This approach enables direct assessment of the imposed displacement field and identification of deviations from ideal behavior, such as wall slip, shear banding, or non-affine deformation.

As a model soft solid, we used a two-component silicone gel (see Section~\ref{sec:samples}) that cures in situ and adheres well to the glass plates. The gel, which exhibits a storage modulus of approximately 1~kPa and negligible flow at small strain, remains optically transparent, making it suitable for imaging. The sample was allowed to cure inside the rheometer for at least three hours prior to testing.

\begin{figure}
    \centering
    \includegraphics[width=0.4\textwidth]{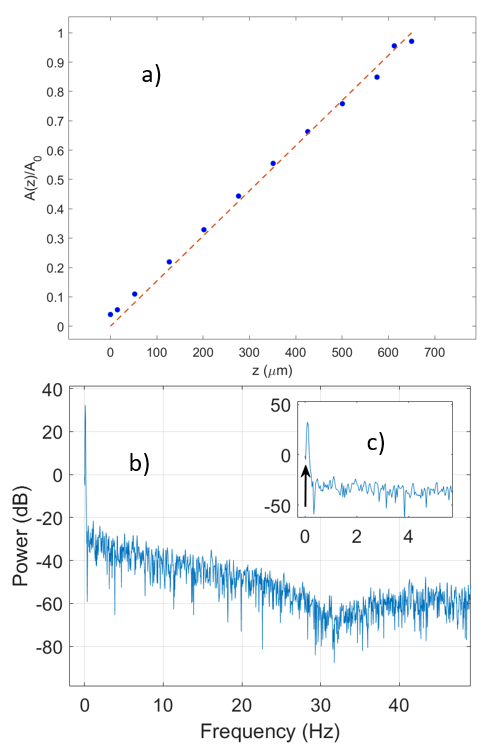}
    \caption{a) Local displacement amplitude normalized by the macroscopic strain amplitude $A(z)/A_0$ as a function of corrected focal position $z$ during oscillatory shear of the soft solid. The linear profile confirms homogeneous shear. Subplots b) and c) show the power spectrum of the stress response; the fundamental frequency is indicated with an arrow in c).}
    \label{fig:profiles_solid}
\end{figure}

We imposed a sinusoidal strain of 5\% amplitude at 1~Hz and acquired bright-field images at 100~fps over 1000 frames. The sample was scanned along the $z$-axis (gap direction) by vertically displacing the objective in steps of 50~\textmu m. Due to refractive index mismatch between the gel and air, the optical position $z'$ must be corrected to yield the actual sample coordinate $z$. The correction is given by $z = z' \cdot \frac{h}{h'}$, where $h$ is the true gap size and $h'$ is the optically measured gap. This correction ensures accurate spatial mapping of the displacement profile. At each focal plane, local frame-to-frame displacements were calculated using the DFT-based image registration algorithm~\cite{guizar2008efficient}. The local displacement amplitude $A_z$ was then normalized by the total macroscopic amplitude $A_0$ obtained from the position sensor, yielding the normalized shear profile $A(z)/A_0$ across the gap. As shown in Fig.~\ref{fig:profiles_solid}, the measured profile is highly linear, confirming that the deformation is transmitted homogeneously across the sample. No evidence of wall slip or shear localization was detected within the experimental resolution.

\begin{figure}
    \centering
    \includegraphics[width=0.4\textwidth]{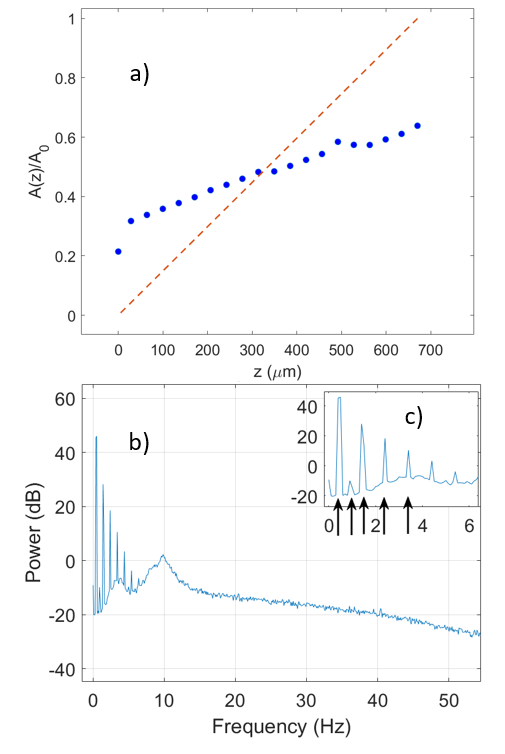}
    \caption{a) Normalized displacement amplitude $A(z)/A_0$ as a function of $z$ across the gap during oscillatory shear of the nanoemulsion. The deviation from the ideal linear profile (red dotted line) is consistent with wall slip. The nearly uniform slope suggests no shear banding. Subplots b) and c) show the power spectrum of the stress response. Arrows in c) indicate the first, second, third, fifth, and seventh harmonics.}
    \label{fig:profiles_nanoemulsion}
\end{figure}

The ability to resolve the local strain field is particularly important for samples prone to shear instabilities. Wall slip and shear banding can significantly affect rheological measurements and complicate material characterization~\cite{becu2005does,bertola2003wall,cloitre2017review}. To demonstrate this capability, we examined a nanometer-scale oil-in-water emulsion (referred to as nanoemulsion), where the small droplet size ensures optical transparency~\cite{chaleshtari2023rheological}. The experimental protocol consisted of applying a sinusoidal strain at 0.5~Hz with 20\% amplitude and scanning the gap using the same procedure described above. The resulting shear profile is shown in Fig.~\ref{fig:profiles_nanoemulsion}. A strong deviation from the ideal linear profile is observed, indicative of pronounced wall slip. The red dotted line shows the ideal linear behavior expected under no-slip conditions. The measured profile has a nearly uniform slope, consistent with wall slip at both boundaries and the absence of internal shear bands. We note that wall slip could have been minimized by using serrated slides; here, smooth slides were used intentionally to allow slip.
For both the soft solid and the nanoemulsion, we analyzed the Fourier spectra of the stress response. In the case of the soft solid, tested within the linear viscoelastic regime, the spectrum exhibits only the fundamental frequency component, as shown in Fig.~\ref{fig:profiles_solid}c. By contrast, the nanoemulsion, subjected to a higher strain amplitude in the nonlinear regime, displays a rich harmonic structure with clear signatures of higher-order harmonics, as shown in Fig.~\ref{fig:profiles_nanoemulsion}c. This harmonic content reflects the nonlinear viscoelastic behavior of the material. Notably, the presence of a contribution corresponding to the second harmonic, in addition to odd harmonics, is consistent with the occurrence of wall slip~\cite{cloitre2017review}.

\subsection{Irreversible rearrangements during LAOS experiments}
Large Amplitude Oscillatory Shear (LAOS) is widely used to probe yielding in soft materials~\cite{ewoldt2008new, rogers2018large, min2014microstructure, poulos2015large}. Frequency-domain analysis of the stress response~\cite{wilhelm1998fourier, wilhelm2002fourier} reveals nonlinear viscoelastic features and provides insight into underlying microscopic processes. Recently, the combination of LAOS with techniques that probe dynamics at different length scales has enabled a more direct connection between macroscopic rheology and microstructural rearrangements.

A particularly powerful method is the \textit{echo protocol}~\cite{laurati2014plastic, petekidis2002rearrangements, petekidis2003yielding, aime2018microscopic}, in which measurements are synchronized to specific phases of the deformation cycle. This stroboscopic approach isolates reversible and irreversible dynamics by comparing configurations at integer multiples of the oscillation period. In microscopy, acquiring images at the same phase of each cycle allows direct visualization of particle trajectories under repeated shear~\cite{aime2019probing_II, villa2022quantitative, edera2024yielding}. Similarly, in scattering experiments, echo-based correlation functions reveal the timescale over which microscopic configurations decorrelate.

\begin{figure}
    \centering
    \includegraphics[width=0.5\textwidth]{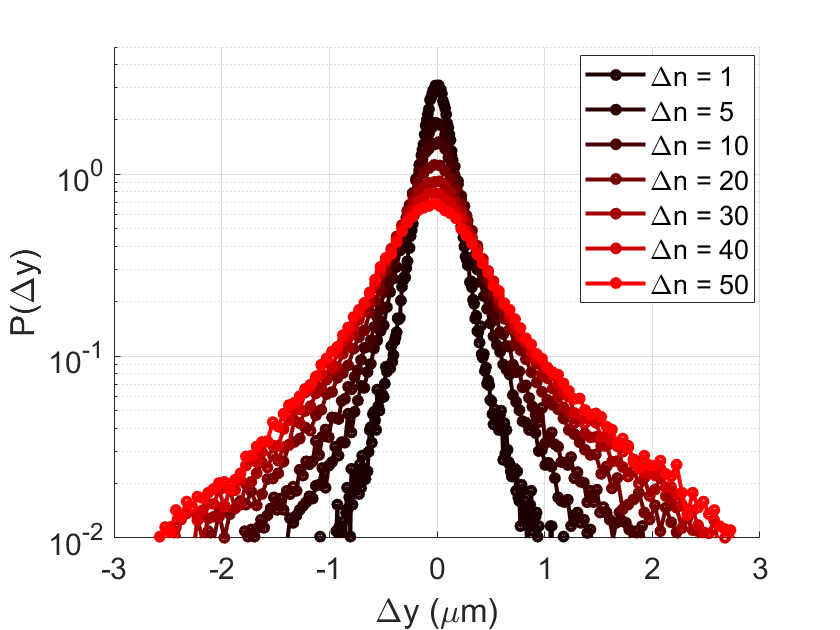}
    \caption{Normalized probability density functions (PDFs) of tracer displacements in the vorticity direction at 45\% strain amplitude (0.29 Hz), as a function of lag period $\Delta n$. Data shown are from a single echo. The progressive broadening indicates accumulation of irreversible rearrangements.} 
    \label{fig:PDF_45}
\end{figure}

\begin{figure}
    \centering
    \includegraphics[width=0.5\textwidth]{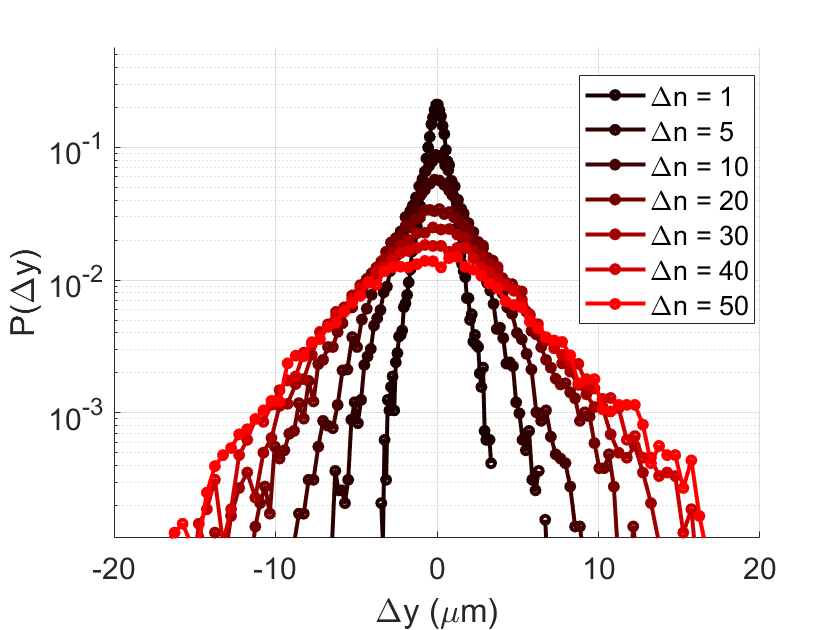}
    \caption{Normalized probability density functions (PDFs) of tracer displacements in the vorticity direction at 80\% strain amplitude (0.29 Hz), as a function of lag period $\Delta n$. Data shown are from a single echo. The broader distributions indicate more extensive irreversible dynamics.}
    \label{fig:PDF_80}
\end{figure}

We demonstrate this protocol using ShearView operating in strain-controlled mode, coupled with phase contrast microscopy. As a model system, we use an overjammed Carbopol suspension (2~wt\% in propylene glycol) seeded with 2~\textmu m polystyrene tracer particles. The sample is imaged at the center of the gap with a 20\texttimes{} objective. Prior to imaging, a pre-shear protocol is applied to erase memory effects: the sample is sheared at 200\% strain amplitude for 100 cycles, with a linear decay of amplitude to the target value. Microscopy images are then acquired stroboscopically at fixed phase using the echo protocol. Two different strain amplitudes are investigated: 45\% (near the crossover of $G'$ and $G''$) and 80\% (deep in the fluidized regime where $G'' > G'$). Representative particle trajectories from a single echo are shown in Supplementary Figures~7 and~8. To quantify non-affine rearrangements, we compute the probability density function (PDF) of particle displacements (for a single echo) in the vorticity direction as a function of lag period $\Delta n$. Results for 45\% strain amplitude are shown in Fig.~\ref{fig:PDF_45}. The progressive broadening of the PDFs indicates cumulative irreversible motion. Analogous data for 80\% strain amplitude are shown in Fig.~\ref{fig:PDF_80}, where the increased width and asymmetry reflect enhanced plasticity.

\subsection{Rearrangements upon shear cessation}
To further illustrate the capabilities of our platform in probing transient microstructural dynamics, we implemented shear cessation experiments and analyzed the ensuing relaxation behavior at both macroscopic and microscopic levels. These protocols are particularly informative for yielding materials, where stress relaxation, recoil, and structural memory can drive complex post-shear dynamics.

\begin{figure}
    \centering
    \includegraphics[width=0.5\textwidth]{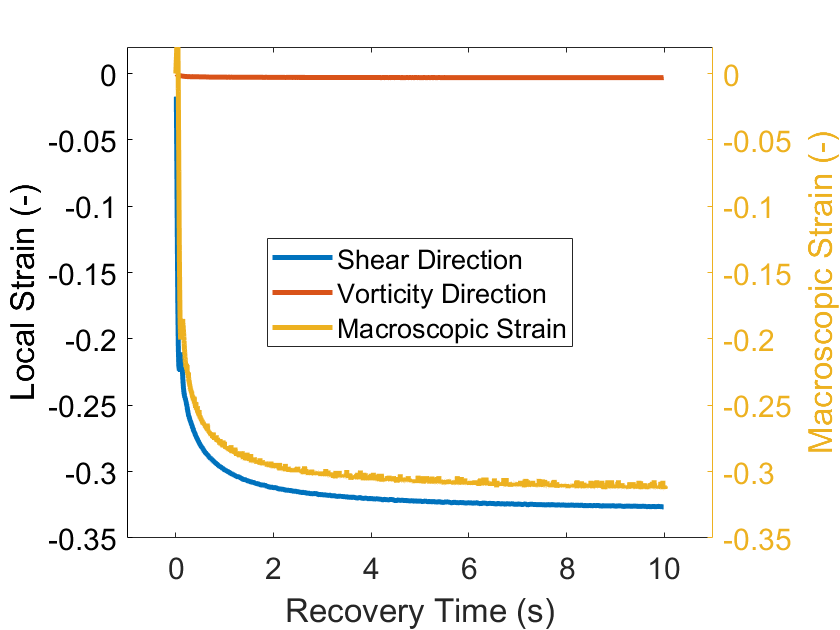}
    \caption{Microscopic strain (blue), obtained via image registration in the field of view, compared with macroscopic strain (yellow), measured by the position sensor, as a function of time after cessation of oscillatory shear. The overall agreement is good, confirming the reliability of the local strain measurement. Minor deviations may arise from uncertainties in determining the exact position of the imaged plane relative to the sample gap, and from the fact that the microscopic strain is evaluated over a small region of interest, which may not fully capture global sample behavior.}
    \label{fig:Macro-microStrain}
\end{figure}

In our experiments, an oscillatory strain with 60\% amplitude at 0.29~Hz was applied to an overjammed Carbopol suspension (2~wt\% in propylene glycol). This strain amplitude lies well beyond the crossover of the storage and loss moduli (see Supplementary Material Fig.~9), placing the system in a fully fluidized state where $G'' > G'$. Oscillatory driving was abruptly stopped at a predefined phase in the cycle, and high-speed imaging (100~fps) was initiated immediately after cessation to capture the material’s recovery dynamics. Microscopic strain was extracted via image registration and compared to the macroscopic strain measured via the position sensor. As shown in Fig.~\ref{fig:Macro-microStrain}, the local and global strains display good agreement throughout the relaxation process. A small discrepancy (approximately 4\% at long times) is attributed to uncertainties in gap calibration and the limited spatial extent of the imaged region, which may not fully capture global behavior.

When performing this type of experiments it is useful that one of the main axes of the camera sensors is parallel to the shear direction. To assess alignment between the imaging axis and the shear direction, we translated a fixed pattern across the field of view and averaged several frames to obtain an image with several streaks oriented along the shear direction. This average image is FFTed and the orientation of the resulting pattern is examined in Fourier space. This procedure typically ensures the angular alignment between the shear direction and the camera axes remains below 1° (see also Supplementary Fig.~10).

To resolve the microscopic dynamics of recovery, we tracked the motion of embedded tracer particles using particle tracking. Given the non-reversible nature of recovery, particle displacements were calculated relative to the final frame of each sequence, where macroscopic strain had fully relaxed. We define these displacements as $\Delta x(t_{\mathrm{end}} - t_i)$, where $t_i$ is the time of the $i^{\text{th}}$ frame and $t_{\mathrm{end}}$ is the final frame. This choice isolates the irreversible component of the relaxation process.

Due to rapid recoil immediately after cessation, reliable tracking was only possible starting around 100~ms after shear was stopped (i.e., from frame 10 onward). We analyzed 14 consecutive frames, corresponding to the main recovery window. A representative trajectory map for a single experiment is shown in Supplementary Material Fig.~11.

\begin{figure}
    \centering
    \includegraphics[width=0.5\textwidth]{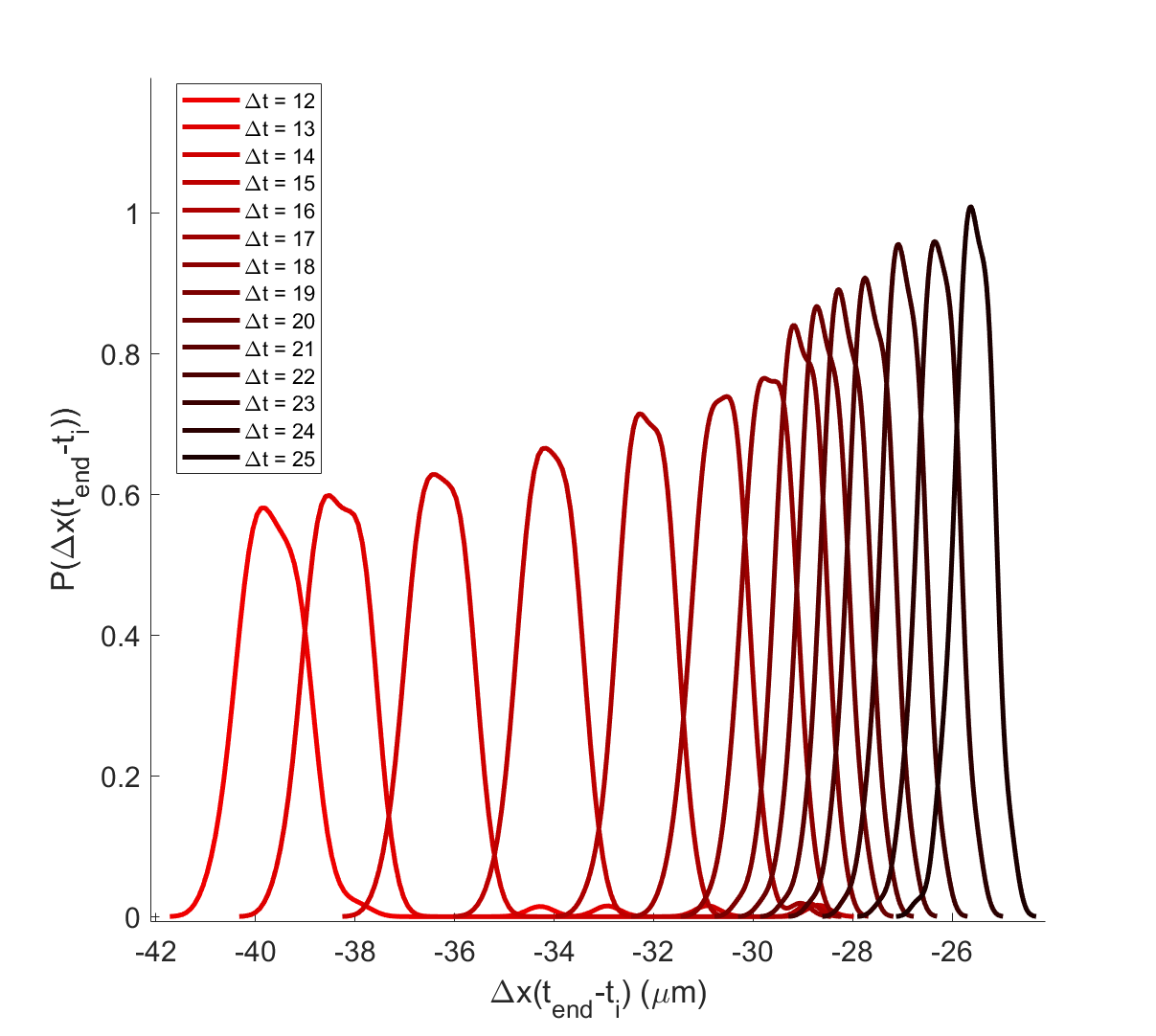}
    \caption{Time evolution of the probability density function (PDF) of tracer particle displacements in the shear direction following cessation of oscillatory shear at 60\% strain amplitude. Displacements are computed relative to the final frame of the sequence. The progressive shift in the distribution reflects ongoing local strain recovery.}
    \label{fig:PDFDx}
\end{figure}

\begin{figure}
    \centering
    \includegraphics[width=0.55\textwidth]{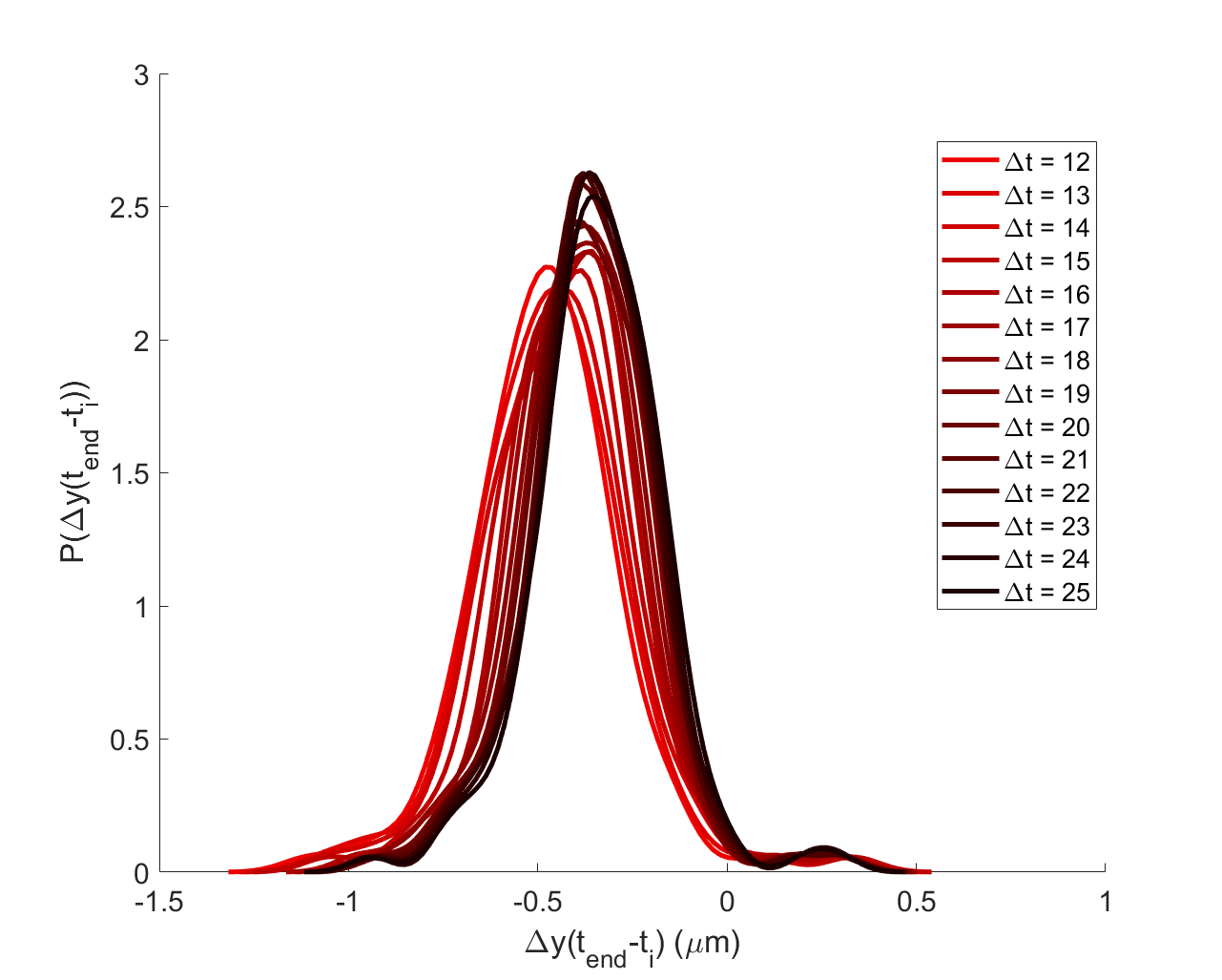}
    \caption{Time evolution of the probability density function (PDF) of tracer particle displacements in the vorticity direction for the same time intervals as in Fig.~\ref{fig:PDFDx}. The distributions remain narrow and centered, indicating minimal transverse motion and confirming that recovery is primarily along the shear direction.}
    \label{fig:PDFDy}
\end{figure}

The displacement distributions provide further insight into the nature of recovery. Figure~\ref{fig:PDFDx} shows the evolution of the probability density function (PDF) of tracer displacements along the shear direction. The progressive shift of the distribution center reflects continued strain relaxation at the microscopic level accompanied by an overall reduction of microscopic mobility mirrored by the narrowing of the PDFs. In contrast, the PDFs along the vorticity direction (Fig.~\ref{fig:PDFDy}) remain symmetric and narrow, consistent with limited transverse motion and no significant residual flow.

These results highlight the strength of our platform in synchronizing rheological protocols with high-resolution, real-time microscopy. The ability to track recovery dynamics at the particle level—following a well-defined mechanical history—enables mechanistic insight into plastic and elastic contributions in soft materials, which is generally inaccessible with bulk rheometry alone.

\section{Samples\label{sec:samples}}
This section summarizes the materials used to characterize and validate the performance of ShearView across a range of mechanical and imaging conditions. The selected systems exhibit diverse viscoelastic properties and span different application domains, including calibration, homogeneous and heterogeneous deformation, nonlinear strain response, and strain recovery. Each formulation was chosen to highlight specific capabilities of the instrument.

\begin{itemize}
    \item \textbf{Viscoelastic melt.} A high-viscosity polydimethylsiloxane (PDMS) oil (Wacker Silicone Fluid; dynamic viscosity 998~Pa$\cdot$s at room temperature) was used as received. This material exhibits pronounced frequency dependence of both storage and loss moduli within the accessible range and served as a benchmark for validating the current-to-force calibration of the voice coil actuator. Comparative measurements were performed using both the compact rheometer and a commercial instrument (Anton Paar MCR 702e).

    \item \textbf{Polystyrene solution in dioctyl phthalate.} A viscoelastic solution was prepared by dissolving 32~wt\% of polystyrene (Mw = 168~kg/mol, Sigma-Aldrich) in dioctyl phthalate (DOP, Sigma-Aldrich), a high-boiling point plasticizer with negligible evaporation. To facilitate dissolution, hexane was used as a co-solvent and subsequently removed by heating the solution at 50$\,^{\circ}$C for 24 hours. The resulting homogeneous solution was used to explore frequency-dependent viscoelasticity.

    \item \textbf{Two-component silicone gel.} A crosslinking silicone elastomer (DOWSIL CY 52-276, Dow) was used as a model soft solid with purely elastic response. Equal volumes of components A and B were hand-mixed according to the manufacturer's protocol. To introduce optical contrast, 5~$\mu$L of a 10~wt\% aqueous suspension of 2~$\mu$m diameter polystyrene particles (MicroParticles GmbH) was dried, ground into a fine powder, and added to component B. The mixture was centrifuged at 6000~rcf for 10 minutes to remove aggregates, and the supernatant was mixed with component A. After thorough mixing (5 minutes) and degassing (500~rcf, 1 minute), the material was loaded into the rheometer and cured in situ for at least 3 hours. The resulting gel adheres strongly to the glass slides and displays linear elastic behavior with a shear modulus of approximately 1~kPa.

    \item \textbf{Over-jammed aqueous microgel suspension.} A Carbopol 971P NF suspension (Lubrizol) was prepared at 5~wt\% by slowly adding the dry powder to deionized water under vigorous stirring (1800~rpm). After full dispersion, the pH was adjusted to 7 using 10~M NaOH solution progressively, followed by 10 minutes of equilibration, between additions. Note that the volume of NaOH in the suspension is not considered in the concentration of the suspension. The dispersion was then aged for at least 24 hours at room temperature. This highly jammed microgel was used to study yielding behavior and to validate large-amplitude strain sweeps.

    \item \textbf{Over-jammed microgel suspension in propylene glycol.} To mitigate solvent evaporation during extended experiments, a Carbopol 974P (Lubrizol) microgel was prepared in propylene glycol, which has a significantly lower vapor pressure than water. The powder was added at 2~wt\% under continuous stirring (7000~rpm) until full dispersion was achieved. The suspension was then sonicated at 50$\,^{\circ}$C for 6 hours and aged for one week. A NaOH solution in propylene glycol was subsequently added to reach a final NaOH concentration of $3 \times 10^{-4}$~g/mL. This formulation was employed for echo and recovery rheology experiments, where long-term mechanical and optical stability is critical.

    \item \textbf{Nanoemulsion.} A fairly transparent oil-in-water nanoemulsion was prepared following the protocol described in Ref.~\cite{chaleshtari2023rheological}. Initially, a coarse emulsion was formed by adding 25~vol\% of low-viscosity silicone oil ($\sim$5~cP) to an aqueous solution of sodium dodecyl sulfate (SDS, 110~mM). The mixture was stirred at 700~rpm for 5 minutes, and 2~$\mu$m polystyrene tracer particles (0.05~wt\%) were added during this stage. The resulting macroemulsion was sonicated using a Fisherbrand ultrasonic processor (model 120) at 90\% amplitude for 20 minutes, while immersed in a temperature-controlled bath maintained at 10$\,^{\circ}$C. Finally, the nanoemulsion was left to equilibrate at room temperature in a Petri dish, where water gradually evaporated until the target oil volume fraction of 0.35 was reached. This formulation exhibits minimal light scattering and is suitable for optical imaging of shear profiles and instabilities.
\end{itemize}

\section{Conclusions and Outlook}

We have introduced ShearView, a compact, cost-effective, and fully open-source linear rheometer capable of both stress- and strain-controlled oscillatory shear experiments, designed to operate in conjunction with high-resolution optical microscopy. The system has been designed for mechanical simplicity, modularity, and reproducibility, while remaining highly adaptable thanks to its entirely software-defined control architecture developed in LabVIEW. The device relies primarily on commercially available components (non-commercial parts can be machined or assembled using the drawings and instructions available in the Github repository) and features a lightweight moving platform stabilized by a linear air bearing, ensuring low mechanical friction and minimal inertia.

We provided a comprehensive assessment of the instrument design and performance, including validation against a high-end commercial rheometer (Anton Paar MCR 702e). Quantitative agreement was achieved across frequency sweeps and large-amplitude oscillatory shear (LAOS) measurements on viscoelastic benchmark samples, demonstrating both the accuracy and robustness of the device. Inertia corrections tailored to the linear geometry were implemented and found to be relevant only for soft samples subjected to high-frequency actuation. In addition, to showcase the capabilities of ShearView, we implemented two recently proposed advanced rheological tests---frequency-modulated mechanical spectroscopy~\cite{geri2018time} and recovery rheology~\cite{donley2020elucidating}---that represent the frontier of rheological material fingerprinting.

The integration of microscopy with synchronized mechanical stimulation was demonstrated through several experimental protocols. These included the measurement of spatially resolved shear profiles, detection of wall slip and shear localization, and stroboscopic imaging under echo protocols for quantifying reversible and irreversible particle dynamics. Additionally, we demonstrated the ability to capture transient behavior through combined rheological and optical measurements during shear cessation. These capabilities, typically inaccessible with commercial platforms, illustrate the value of ShearView for multi-scale soft matter investigations.

All hardware designs, control software, and example datasets are freely available to ensure reproducibility, encourage community contributions, and support deployment in both research and teaching contexts. In educational settings, ShearView offers hands-on experience with rheometry, control theory, data acquisition, and microscopy, providing a modular and transparent platform for graduate-level training.

Despite its versatility, ShearView retains certain practical limitations. As a research-oriented apparatus, it is best suited to experienced users; unlike commercial rheometers, adjustments such as gap setting, alignment, and inertia determination are manual, which introduces a learning curve for first-time operation. Measurement sequences are not yet fully automated: for example, in a dynamic strain sweep, each strain amplitude must be probed independently, with results saved and analyzed offline. This reflects the need for slightly different PID parameters at different amplitudes, which are presently tuned manually. In practice, once optimal PID parameters were identified for a given geometry and material class, they remained stable across multiple sessions, with only minor (yet needed) adjustments required when switching between samples of markedly different stiffness or viscosity.

The current implementation does not include temperature or environmental control. While integration into a microscope incubator can provide temperature, humidity, and gas regulation, this requires dedicated equipment and typically longer equilibration times than on a commercial rheometer. Finally, the usable range of material properties is constrained: very low-viscosity samples ($<100$~mPa\,s), very high-modulus samples ($>10^5$~Pa), or materials generating large normal forces are not well-suited to the current geometry.

\paragraph*{Outlook and perspective.}
Further refinement of the control system, including migration to LabVIEW Real-Time OS or adoption of specialized hardware such as Moku (Liquid Instruments), could eliminate latencies associated with general-purpose operating systems, improving actuation fidelity and expanding the accessible frequency range. Such hardware could also enable a fully hardware-based feedback loop, increasing robustness and speed. Beyond these technical enhancements, ShearView offers a powerful platform for studying mechanical training, memory formation, and history-dependent behavior in amorphous soft materials such as colloidal glasses, gels, and dense suspensions. These systems exhibit complex mechanical responses under cyclic loading, including strain hardening, softening, aging, and memory encoding—phenomena that require precise, reproducible deformation protocols combined with high-resolution spatial measurements. The combined stress/strain control and real-time imaging capabilities of ShearView make it ideally suited for such investigations, enabling joint interrogation of mechanical inputs and structural rearrangements over extended temporal and spatial scales. We anticipate that future studies using ShearView will contribute to a deeper understanding of how soft disordered materials evolve under repeated mechanical stimulation~\cite{chen2025microstructural}, and how such training protocols might be exploited in applications ranging from soft robotics to adaptive biomaterials.

\begin{acknowledgments}
We gratefully acknowledge insightful discussions and valuable suggestions from Matteo Brizioli, Stefano Aime, Stefano Villa, Athanasios Athanasiou, Luca Cipelletti, Mohan Das, Reza Foudazi, Fabio Giavazzi, Muthukumar M, George Petekidis, Simon Rogers and Véronique Trappe. We thank Lubrizol for kindly supplying the Carbopol (microgel) powder, and Véronique Trappe for providing the Carbopol suspension in propylene glycol. We acknowledge support from the Austrian Science Fund (FWF), Grant DOI 10.55776/PIN9311124 (Transforming Gels Through Training – TRAINGEL).
\end{acknowledgments}

\section{Data Availability Statement}
All hardware design files, control software, and analysis scripts associated with this study are openly available to facilitate reproducibility and reuse. The following repositories and licenses apply:

\begin{itemize}
  \item \textbf{Instrument control} (LabVIEW scripts) and \textbf{rheological analysis} (MATLAB scripts) \textbf{software}:
  Available at \faGithub{} \url{https://github.com/somexlab/shearview} under the \textbf{MIT license}.
  
  \item \textbf{Hardware design files} (CAD models, bill of materials, wiring diagrams, and mechanical drawings):
  Available at \faGithub{} \url{https://github.com/somexlab/shearview} under the \textbf{CERN Open Hardware License Version 2 (CERN-OHL-P v2)}.
  
  \item \textbf{Example datasets} used in the figures of this manuscript are archived on \textbf{Phaidra (University of Vienna repository)}: \url{https://doi.org/10.25365/phaidra.706}
\end{itemize}

\section{Supplementary Material}

Supplementary figures referenced in the main text are available in the Supplementary Material.

\bibliography{kalafatakis}

\end{document}



\title[ShearView - Supplementary Material]{ShearView: A Compact Stress- and Strain-Controlled Rheometer for Integrated Rheo-microscopy (Supplementary material)}


\author{Nikolaos Kalafatakis}
\author{Roberto Cerbino}%
\affiliation{ 
Faculty of Physics, University of Vienna, Boltzmanngasse 5, 1090, Vienna, Austria} 
\date{\today}


\pacs{}

\maketitle 


\begin{figure}
    \includegraphics[scale=0.6]{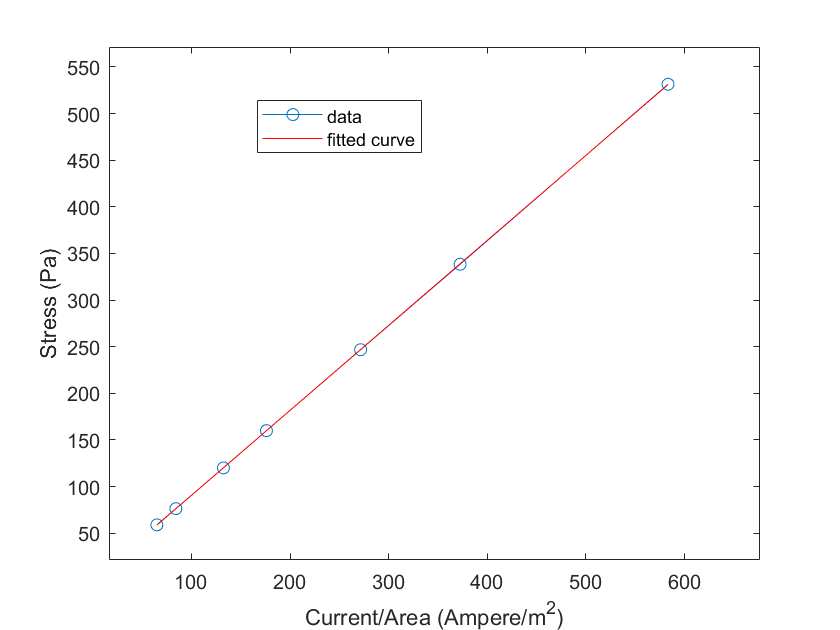}
    \caption{Calibration of the force constant of the linear actuator. The complex modulus in the linear viscoelastic regime of the high-viscosity PDMS melt is measured using the MCR 702e rheometer. The same strain amplitudes are then applied and measured with our compact rheometer. Using the complex modulus, the corresponding stresses for the strain amplitudes measured with the compact rheometer are calculated. The amplitudes of current required to achieve these strain amplitudes are normalized by the sample area. The slope of the stress versus normalized current curve defines the force constant of the actuator. The resulting force constant is: $F = 0.911 \pm 0.002$ \textit{N/A}. }
    \label{fig:force}
\end{figure} 

\begin{figure}
    \includegraphics[scale=0.5]{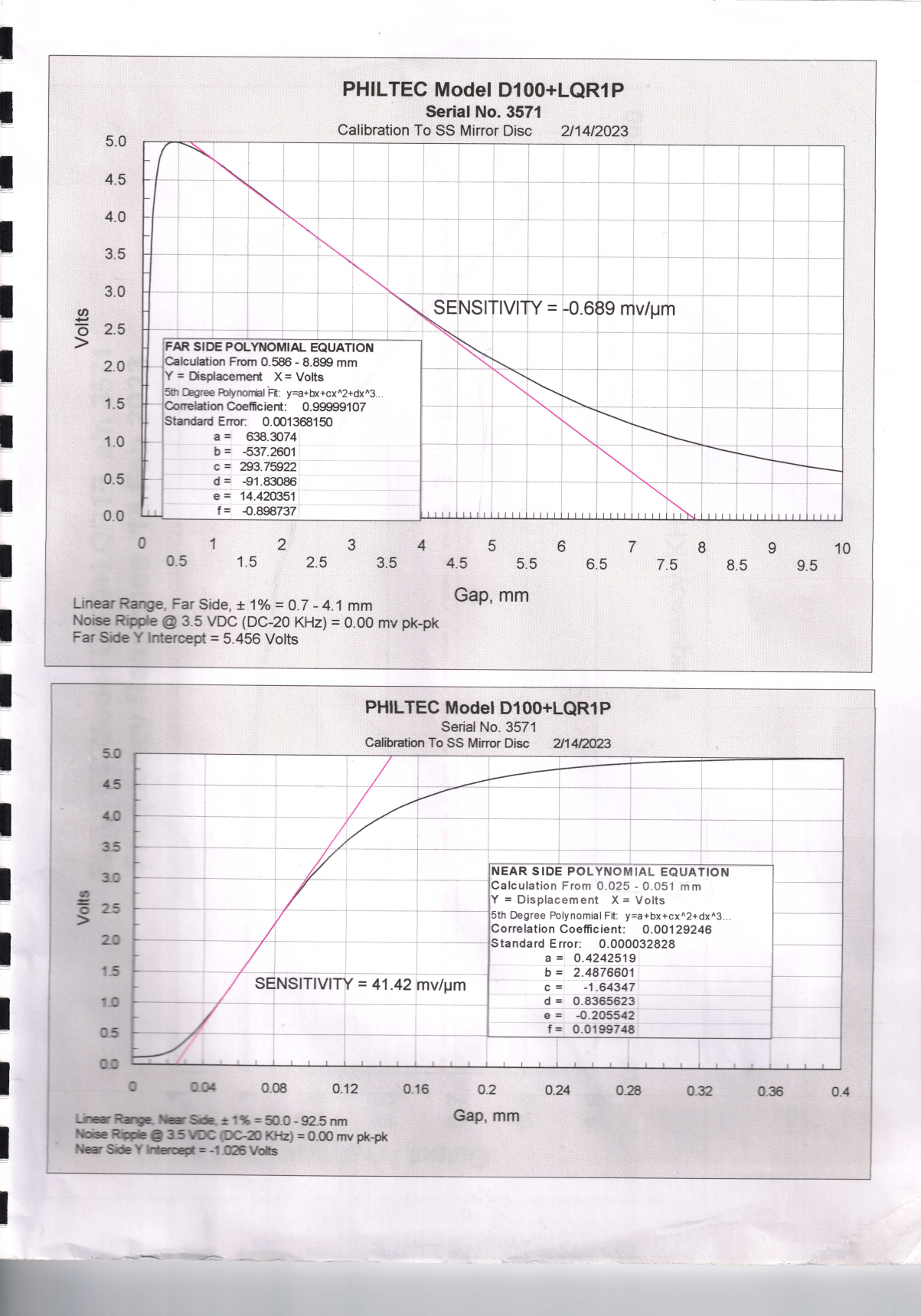}
    \caption{Calibration data provided from the displacement sensor manufacturer.}
    \label{fig:philtec}
\end{figure} 

\begin{figure}
    \includegraphics[scale=0.4]{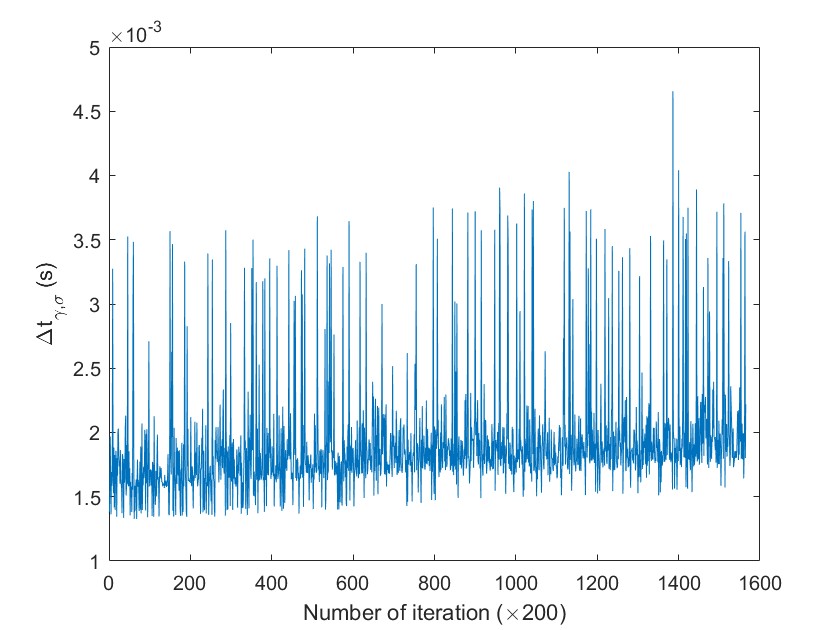}
    \caption{Time difference between measurement of the displacement and application of current to the actuator.}
    \label{fig:time}
\end{figure}

\begin{figure}
    \includegraphics[scale=0.3]{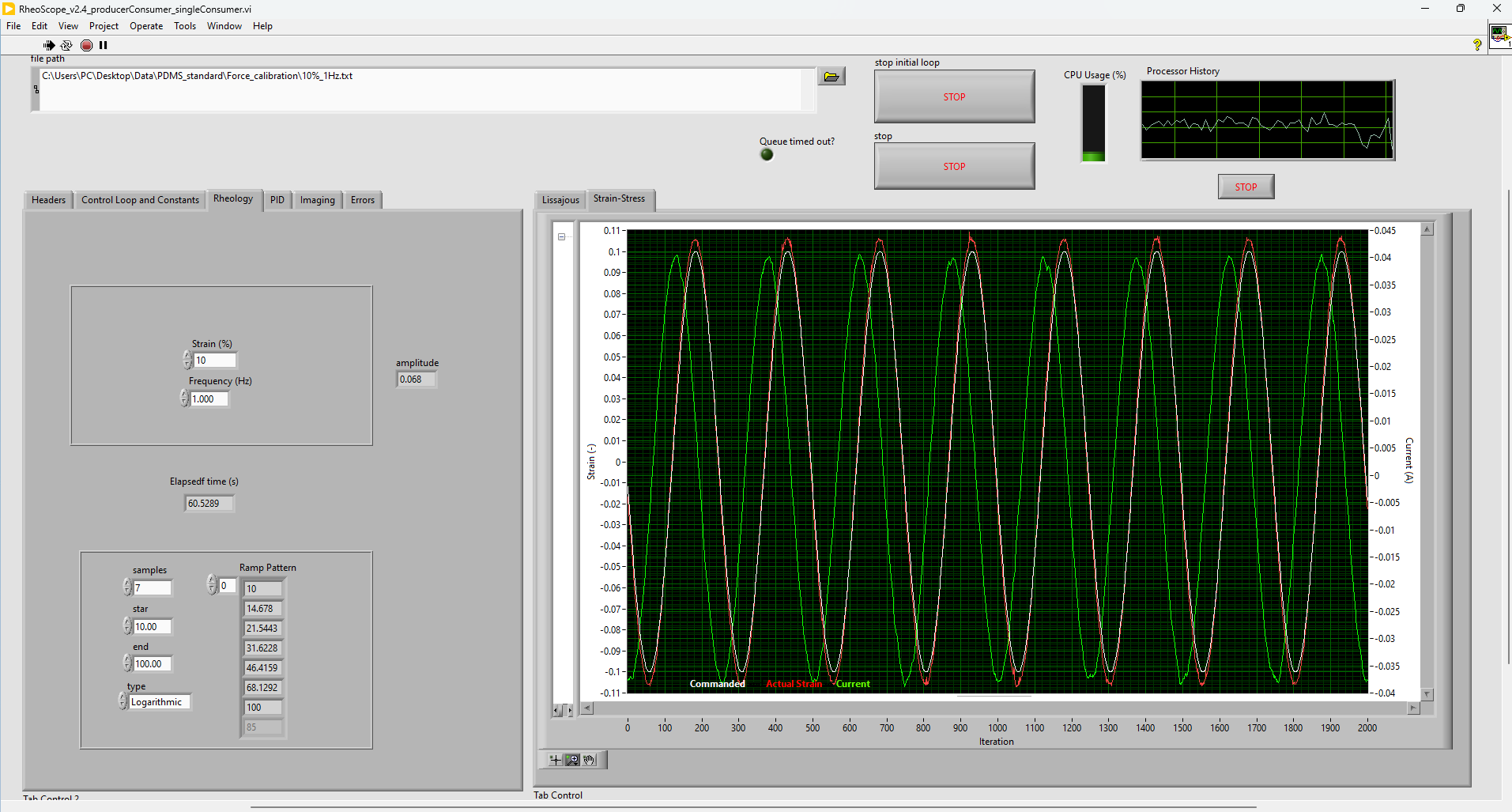}
    \caption{LabVIEW front-panel (user interface) of one of the vi's (virtual instrument). The interface allows file path selection for data saving, header selection, parameter definition (e.g. gap, rheology parameters such as frequency and strain, PID parameters, imaging parameters, etc.), error reporting and live plotting of the strain and current as a function of time, or as Lissajous plot.}
    \label{fig:LabVIEW}
\end{figure}

\begin{figure}
    \includegraphics[scale=0.5]{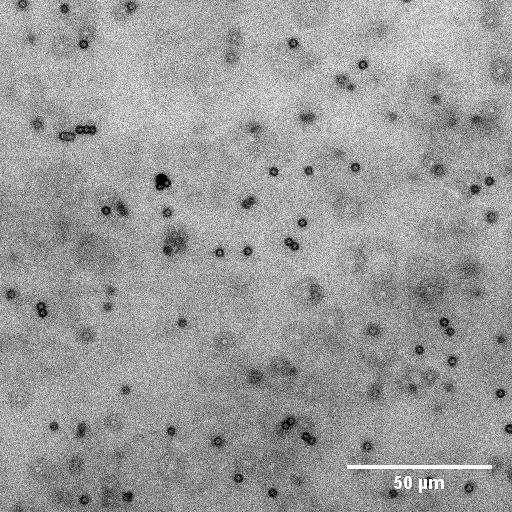}
    \caption{Representative image acquired in bright field with the same parameters as in the image acquired in phase contrast shown in the main text. Sample: overjammed Carbopol suspension (2 wt\% in propylene glycol).}
    \label{fig:bright}
\end{figure}

\begin{figure}[!tbp]
    \centering
    \subfloat[Fringes observed for non-aligned microscope glasses]{\includegraphics[width=0.45\textwidth]{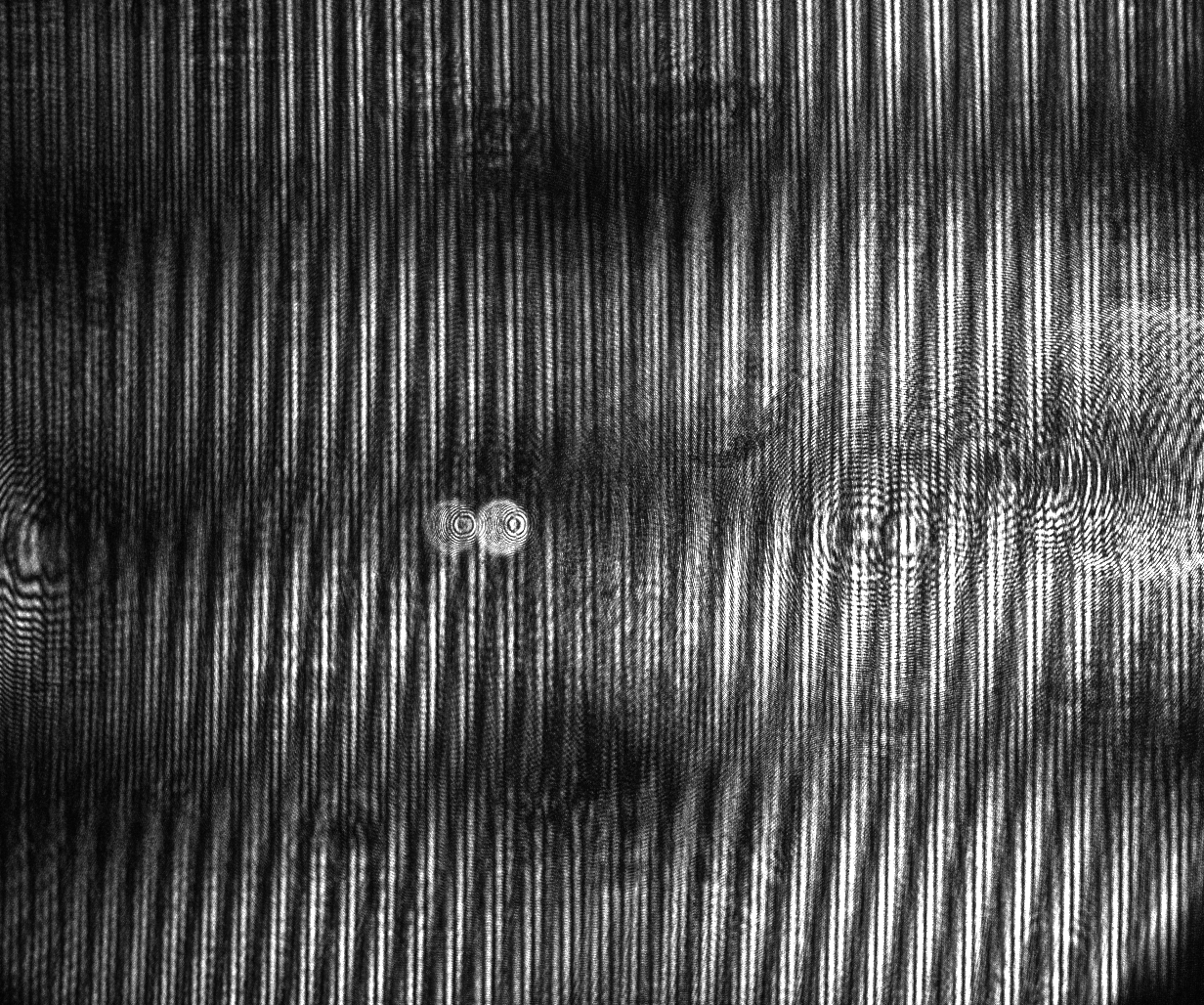}\label{fig:image1}}
    \hfill
    \subfloat[Fringes observed when the parallelism of the glasses is at its maximum.]{\includegraphics[width=0.45\textwidth]{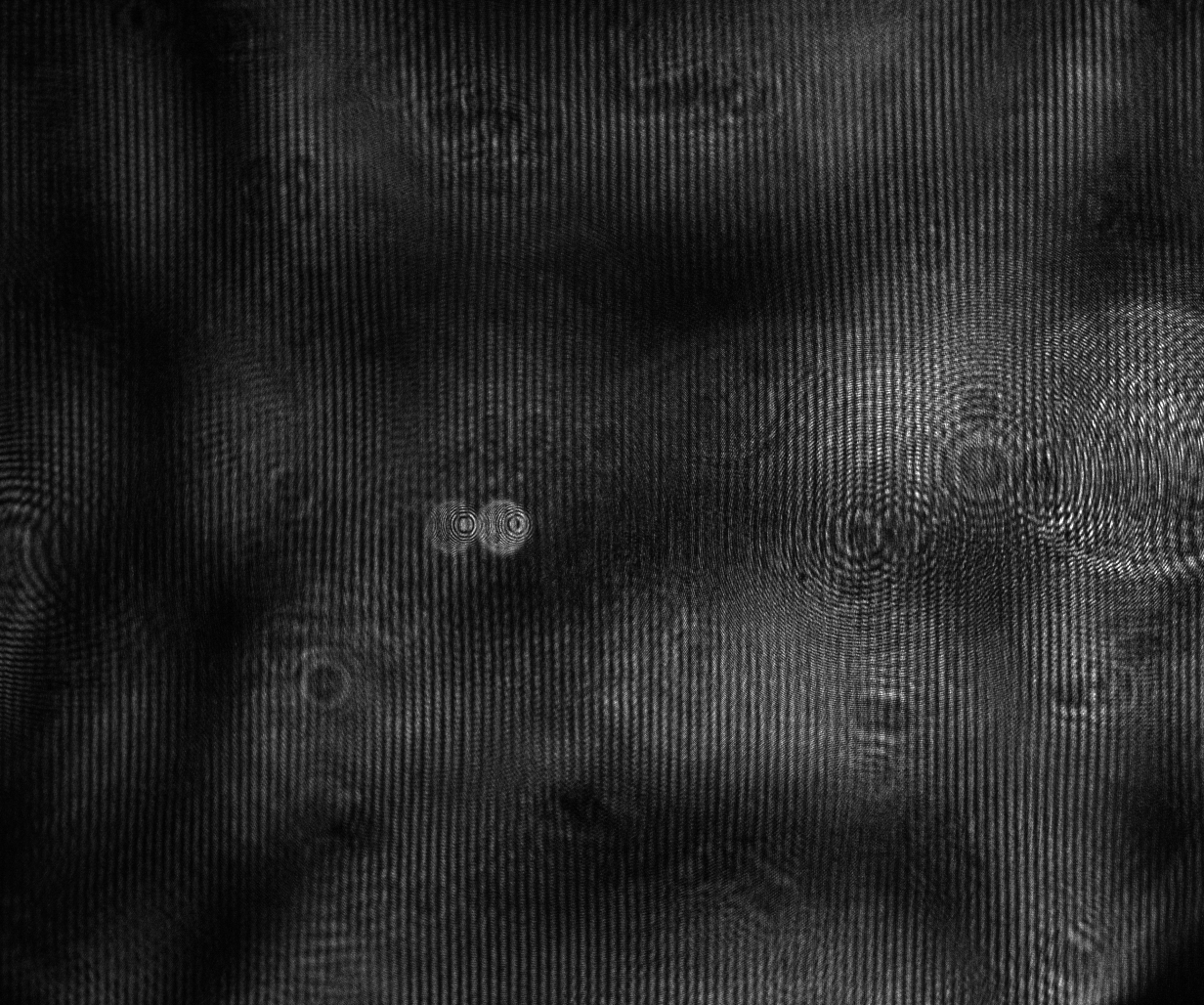}\label{fig:image2}}
    \caption{The fringes observed (a) before and (b) after alignment of the parallelism of the two microscope glasses. Parallelism is optimized when interference fringes are minimized; shallow gradients as seen in (b) indicate good alignment.}
\end{figure}

\begin{figure}
    \centering
    \includegraphics[width=0.9\textwidth]{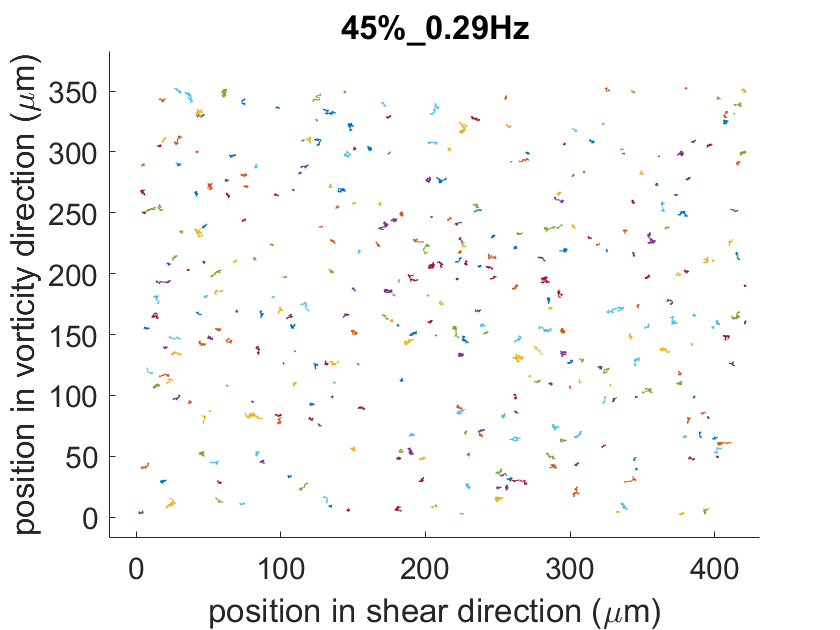}
    \caption{Particles trajectories of a stroboscopic sinusoidal oscillatory shear at strain amplitude of 45\% and frequency of 0.29 Hz. Sample: overjammed Carbopol suspension (2 wt\% in propylene glycol). }
    \label{fig:echoes_45}
\end{figure}

\begin{figure}
    \centering
    \includegraphics[width=0.9\textwidth]{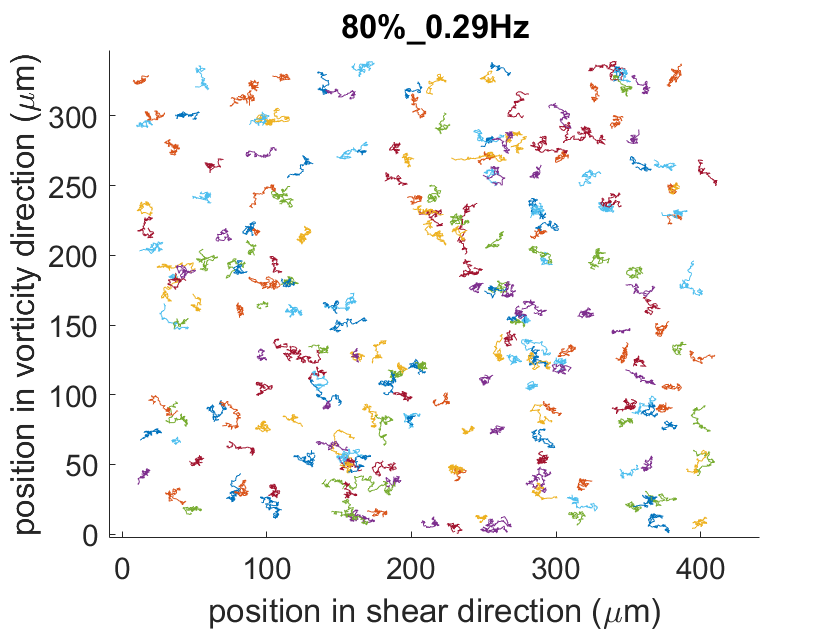}
    \caption{Particles trajectories of a stroboscopic sinusoidal oscillatory shear at strain amplitude of 80\% and frequency of 0.29 Hz. Sample: overjammed Carbopol suspension (2 wt\% in propylene glycol).}
    \label{fig:echoes_80}
\end{figure}

\begin{figure}
    \centering
    \includegraphics[scale=0.4]{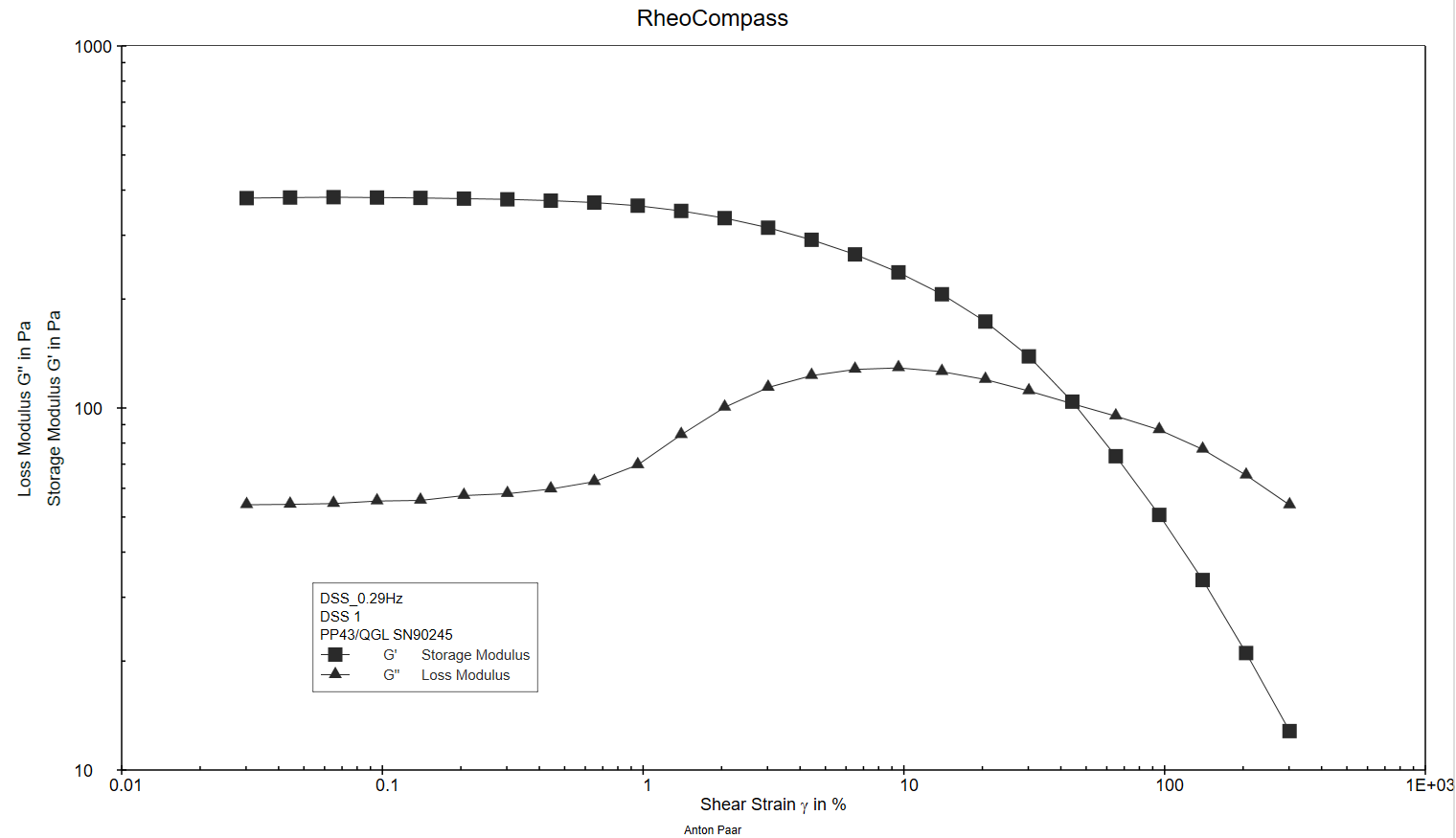}
    \caption{Strain sweep of the overjammed Carbopol suspension (2 wt\% in propylene glycol) at 0.29 Hz.}
    \label{fig:Carbopol}
\end{figure}

\begin{figure}[!tbp]
    \centering
   {\includegraphics[width=0.5\textwidth]{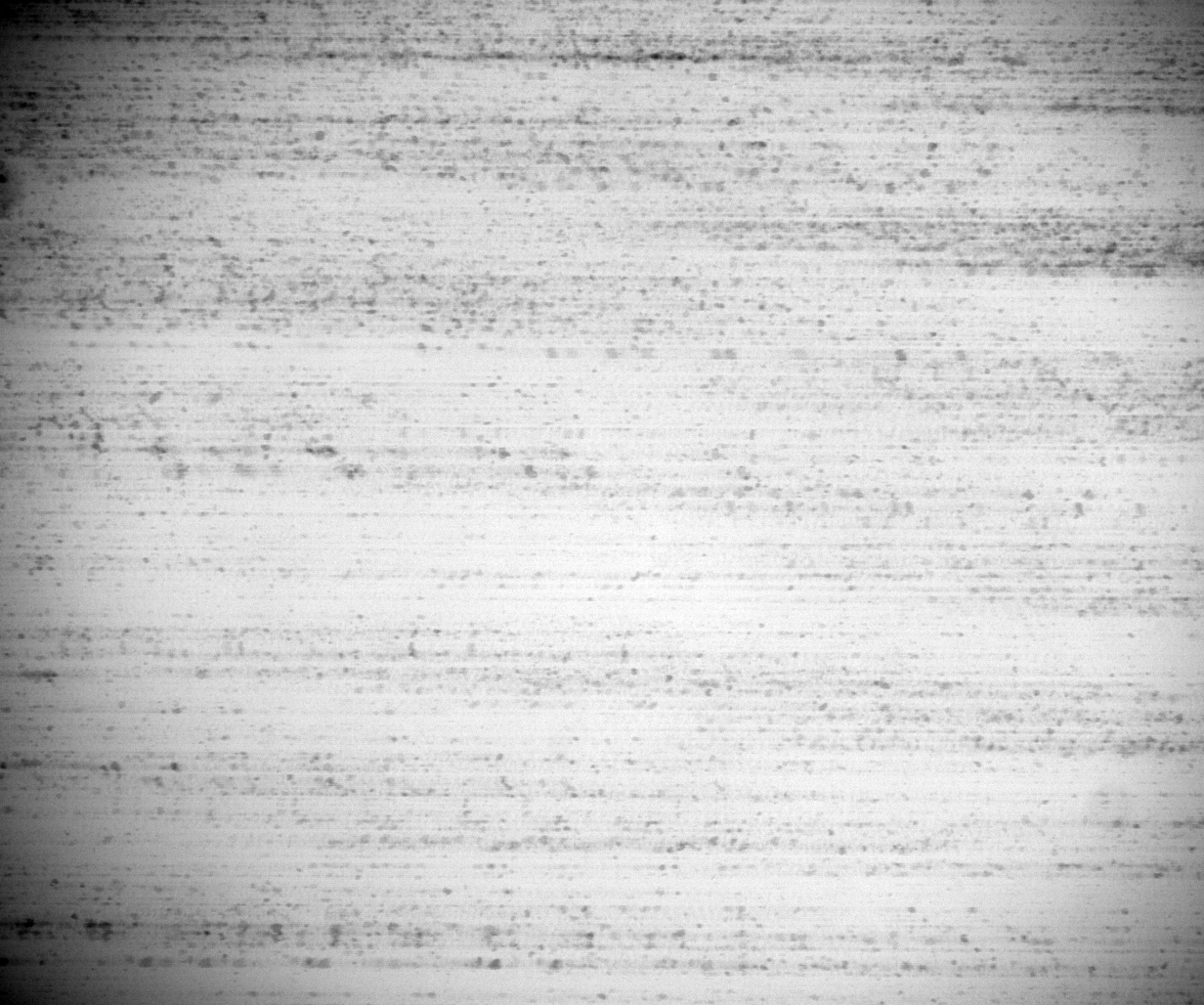}}\label{fig:image1}
    \hfill
    {\includegraphics[width=0.45\textwidth]{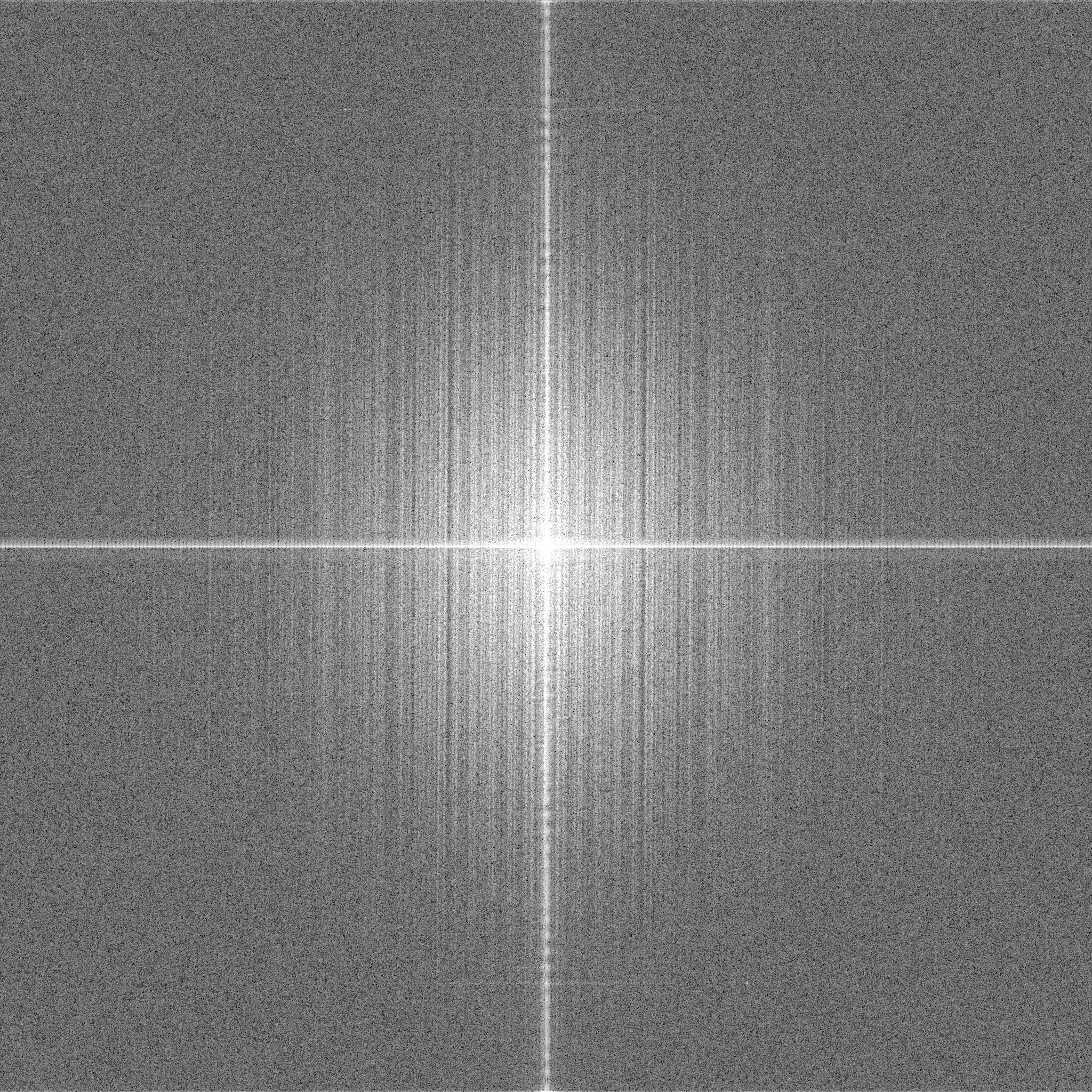}\label{fig:image2}}
    \caption{(Left) Average image obtained by superimposing images of a fixed pattern translated along the shear direction, producing elongated streaks aligned with shear. (Right) 2D Fourier transform (FFT) of the average image showing a cross pattern—arising from residual intensity inhomogeneity—modulated by the diffraction pattern generated by the horizontal streaks, which extends predominantly along the vertical direction. If the streaks were not horizontal, the vertical spoke would be tilted accordingly. In this case, the sensor’s horizontal axis is aligned with the shear direction to within $1^{\circ}$.}
\end{figure}

\begin{figure}
    \centering
    \includegraphics[width=0.9\textwidth]{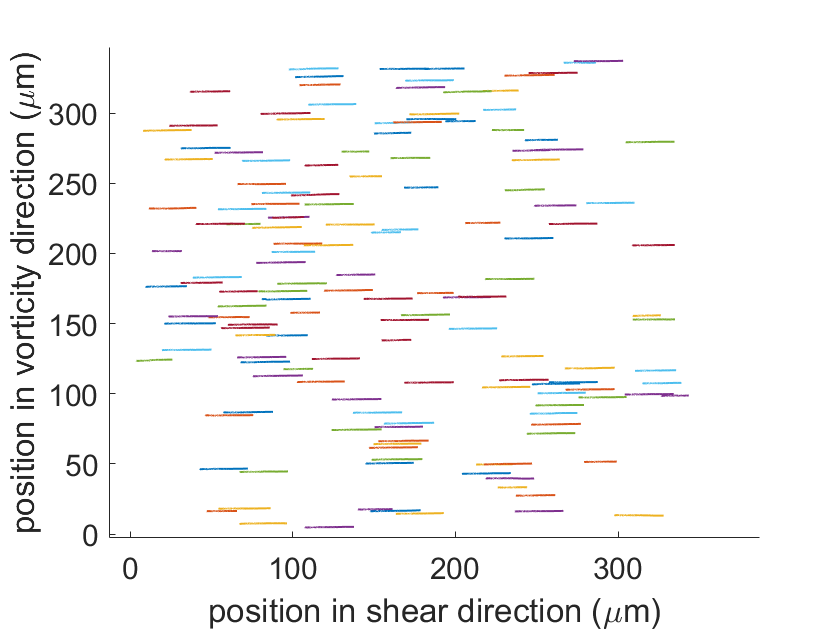}
    \caption{Representative trajectories of particles reproduced from particle tracking upon shear cessation. A small area of the field of view is shown here. Sample: overjammed Carbopol suspension (2 wt\% in propylene glycol).} 
    \label{fig:trajectoriesRecovery}
\end{figure}

\begin{figure}
    \centering
    \includegraphics[scale=0.5]{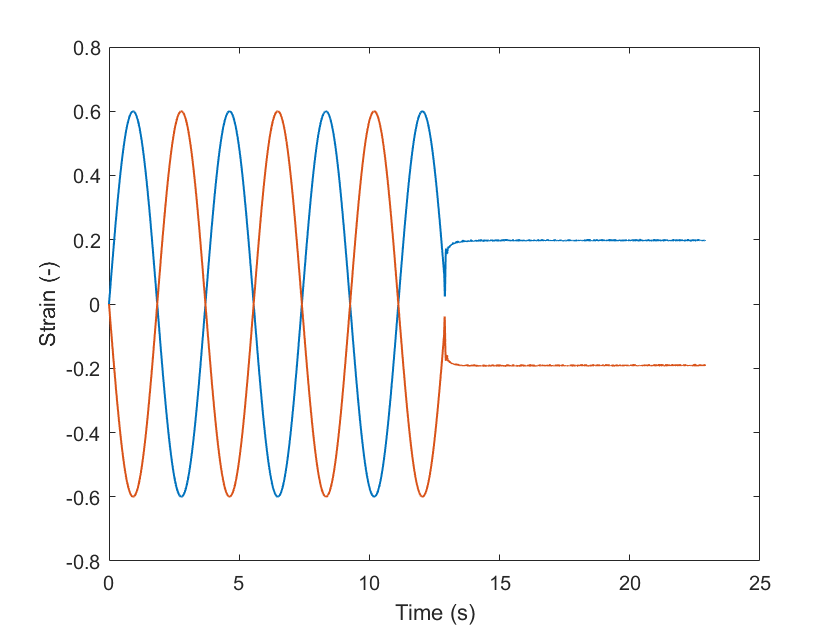}
    \caption{Representative data acquired in recovery rheology. Strain as a function of time acquired at the "forward" and "reverse" direction. Sample: overjammed Carbopol suspension (2 wt\% in propylene glycol).}
    \label{fig:rec2}
\end{figure}


%
%

%

